\def\draftversion{false}

\RequirePackage{ifthen}
\ifthenelse{\equal{\draftversion}{true}}{
  \documentclass[prb,galley,showpacs,preprintnumbers,citeautoscript,
      amsmath,amssymb,longbibliography]{revtex4-1}
}{
  \documentclass[citeautoscript,floatfix,aps,prx,twocolumn,
      superscriptaddress,longbibliography]{revtex4-1}
}

\usepackage{amsmath}
\usepackage{graphicx}
\usepackage{epstopdf}
\usepackage{natbib}
\usepackage{array}
\usepackage{dcolumn}
\usepackage{bm}

\usepackage{soul}  

\usepackage[usenames,dvipsnames]{color}

\soulregister\cite7

\ifthenelse{\equal{\draftversion}{true}}{
  \usepackage{showlabels}
  \marginparwidth 2.7in
  \marginparsep 0.5in
  \newcounter{comm} 
  \def\commnext{\stepcounter{comm}}
  \def\commtext{{\bf\color{blue}[\arabic{comm}]}}
  \def\commmar{{\bf\color{blue}[\arabic{comm}]}}
  \def\dvm#1{\commnext\marginpar{\small DV\commmar: #1}\commtext}
  \def\cdm#1{\commnext\marginpar{\small CED\commmar: #1}\commtext}
  \def\msm#1{\commnext\marginpar{\small MS\commmar: #1}\commtext}
  \def\asm#1{\commnext\marginpar{\small AS\commmar: #1}\commtext}
  \def\miq#1{\commnext\marginpar{\small MR\commmar: #1}\commtext}
  \def\mlab#1{\marginpar{\small\bf #1}}
  
}{
  \def\dvm#1{}
  \def\cdm#1{}
  \def\msm#1{}
  \def\asm#1{}
  \def\miq#1{}
  \def\mlab#1{}
  
}

\def\nn{\nonumber\\}

\def\E{\mathcal{E}}
\def\EE{\bm{\E}}
\def\Ea{\mathcal{E}_\alpha}
\def\Eb{\mathcal{E}_\beta}

\begin{document}

\title{First-principles theory of spatial dispersion: 
 Dynamical quadrupoles and flexoelectricity}

\author{Miquel Royo}
\affiliation{Institut de Ci\`encia de Materials de Barcelona 
(ICMAB-CSIC), Campus UAB, 08193 Bellaterra, Spain}

\author{Massimiliano Stengel}
\affiliation{Institut de Ci\`encia de Materials de Barcelona 
(ICMAB-CSIC), Campus UAB, 08193 Bellaterra, Spain}
\affiliation{ICREA - Instituci\'o Catalana de Recerca i Estudis Avan\c{c}ats, 08010 Barcelona, Spain}

\date{\today}

\begin{abstract} 
Density-functional perturbation theory (DFPT) is nowadays the method of choice for the 
accurate computation of linear and non-linear response properties of materials from first principles.
A notable advantage of DFPT over alternative approaches is the possibility of
treating incommensurate lattice distortions with an arbitrary wavevector, ${\bf q}$, at essentially 
the same computational cost as the lattice-periodic case.
Here we show that ${\bf q}$ can be formally treated as a perturbation 
parameter, and used in conjunction with established results of perturbation theory 
(e.g. the ``$2n+1$'' theorem) to perform a long-wave expansion of an arbitrary response 
function in powers of the wavevector components. 
This provides a powerful, general framework to accessing
a wide range of spatial dispersion effects that were formerly 
difficult to calculate by means of first-principles electronic-structure methods.
In particular, the physical response to the spatial gradient of any
external field can now be calculated at negligible cost, by using the response 
functions to \emph{uniform} perturbations (electric, magnetic or strain fields) as the 
only input. 
We demonstrate our method by calculating the flexoelectric and dynamical quadrupole
tensors of selected crystalline insulators and model systems. 
\end{abstract}

\maketitle

\section{Introduction}

Spatial dispersion refers to a dependence of a
material property (e.g. permittivity, conductivity or 
phonon frequency) on the wavevector ${\bf q}$ at which 
it is probed, or equivalently on the gradients of
the perturbation and/or the response in real space.
Its origin can be traced back to the nonlocality of
the microscopic interactions in condensed-matter systems, 
where the response to an external field (electromagnetic
field or atomic displacement) typically occurs over a 
neighborhood of the point where the field is applied.
While in general such effects are weak and 
can often be neglected in macroscopic theories, there
are several instances where their physical consequences are 
important, both regarding their fundamental interest and their
potential towards practical applications.
Indeed, with the ongoing interest in nanoscale phenomena,
researchers are increasingly often facing situations where 
the relevant scalar, vector or tensor quantities (e.g. polarization or strain) 
display large variations on a very small length scale; 
this is precisely the regime at which gradient effects
can become strong.

Historically, spatial dispersion has been most studied in 
the context of the optical response. The first-order wavevector 
dependence of the dielectric susceptibility tensor, for example, 
is responsible for the \emph{natural optical activity},~\cite{Malashevich-10,electrotoroidic}
which is the property of some crystals of rotating the plane of 
polarization of the transmitted light. 
Manifestations of spatial dispersion are, however, ubiquitous;
they can involve magnetism (the magnetoelectric effect can be regarded as
the first-order dispersion of the conductivity~\cite{Malashevich-10}) or elastic 
degrees of freedom as well (the counterpart of optical gyrotropy
in phononics is known as \emph{acoustical activity}~\cite{Portigal-68}).
In the latter context, flexoelectricity~\cite{tagantsev} is arguably the most 
notable example, as it has been intensely explored both
experimentally and theoretically in the past ten years or so.~\cite{pavlo_review,chapter-15}
It describes the polarization response to the \emph{gradient} 
of the applied strain, and therefore it can be understood as the 
spatial dispersion of the piezoelectric tensor.
Being it a universal property of all insulators regardless of crystal
symmetry, it provides a tantalizing route to novel electromechanical 
device concepts,~\cite{Bhaskar2016} and opens the way to many other applications in
energy and information technology.~\cite{Lu2012,Narvaez16}

The long-wavelength regime also occupies a central place in the theory 
of lattice dynamics in insulators.
Indeed, in the ${\bf q}\rightarrow 0$ limit, phonons in insulating crystals 
are associated with macroscopic electric fields that are due to the long-range
electrostatic interactions between atoms.~\cite{born/huang}
Identifying and correctly treating such long-range contributions is
crucial for a meaningful calculation of the interatomic force constants (IFC)
from first principles.~\cite{Gonze/Lee,Baroni-01}
The former are typically written as electrostatic dipole-dipole terms, which 
are responsible for the well-known 
frequency splitting between longitudinal (LO) and transverse (TO) optical phonon 
branches.
It is important to note, however, that the dipole-dipole term only captures the 
\emph{leading} contribution to the long-range IFCs. 
(Dipole-dipole terms decay as $1/d^3$ as a function of the interatomic distance, $d$.)  
Higher orders ($1/d^4$ and faster) are always present, but are systematically neglected
as their physical consequences are much more subtle. 

The next lowest order, for example, involves dipole-quadrupole interactions
and is responsible for a nonanalytic behavior~\cite{artlin} of the force-constant matrix
at $\mathcal{O}(q^1)$; this translates into a $1/d^4$ decay in real space of the
corresponding contribution to the IFCs.
A quadrupolar response to an atomic displacement requires a broken-inversion symmetry 
environment to be active, and is always (although not exclusively) present in piezoelectric 
crystals. Interestingly, in his seminal 1972 paper, R.M. Martin predicted~\cite{Martin} that the 
electronic contribution to the piezoelectric tensor can be written as a sublattice sum of 
the ``dynamical quadrupoles'', so we expect these couplings to be important in compounds 
where electromechanical effects are strong.
However, viable methods to compute the quadrupole tensor have been lacking to date;
this quantity can be regarded as the first-order spatial dispersion of the Born effective charge 
tensor, and is therefore characterized by analogous technical challenges as the calculation of the
flexoelectric tensor.

Developing a systematic, quantitative theory of such effects would be 
very desirable to improve their fundamental understanding and support 
the ongoing experimental efforts.
Achieving this goal, however, presents considerable technical difficulties 
from the point of view of first-principles electronic-structure 
theory, due to the inherent breakdown of 
translational periodicity that a spatial gradient entails.
In the case of flexoelectricity,\footnote{
  We discuss this specific example here, as flexoelectricity is a paradigmatic
  case of dispersion effect where the theoretical efforts have been 
  most successful in the recent past. The following considerations, however, 
  qualitatively hold for the entire class of physical properties that 
  we mentioned in the previous paragraph.}
for example, several routes have been explored to cope with this issue.
Initially, the flexoelectric coefficients were written as 
real-space moments of the response (either the electronic
charge density or the atomic forces) to the displacement of
an isolated atom~\cite{Hong-11,Hong-13}. 
Later, the real-space sums were recast as small-${\bf q}$ expansions
of the response to a monochromatic displacement pattern at
a given wavevector ${\bf q}$.~\cite{artlin,artgr,artcalc}
Other subtleties were addressed as well, such
as the definition and implementation of the current-density
response,~\cite{Cyrus} which eventually allowed for the calculation of the
bulk flexoelectric tensor within a perturbative
framework based on a primitive cell of the crystal.~\cite{Cyrus}

While the strategy of Ref.~\onlinecite{Cyrus} 
could be, in principle, generalized to other physical 
properties, it still presents an important drawback.
Several linear-response calculations need to be performed at 
different ${\bf q}$-points in a vicinity of the Brillouin zone 
center, and the second-order coefficients (corresponding to the 
flexoelectric tensor components) are then extracted via a numerical fit.
This introduces a significant computational overhead (to
repeat the same calculations at several values of ${\bf q}$),
and is a potential source of numerical inaccuracies related 
to the fit.
It would be much cheaper from the computational point of view, and 
convenient from the point of view of the end user, to \emph{directly} 
calculate the desired dispersion coefficients as part of the intrinsic 
linear-response capabilities of the code. 
To achieve this goal, however, one needs first to establish 
a general formalism to describing the long-wavelength limit 
within the context of DFPT.

Here we provide a comprehensive solution to the above issues by
first rewriting the second-order energy at finite ${\bf q}$ as
an \emph{unconstrained} minimization problem of a variational
functional of the first-order wavefunctions.
Next, we show that the parametric ${\bf q}$-dependence of the second-order 
energy can be regarded as a small perturbation of the ${\bf q}=0$ functional;
hence, one can apply the standard tools of DFPT to perform an analytic 
long-wavelength expansion of an arbitrary response property of the crystal
in powers of ${\bf q}$.
Remarkably this strategy, in combination with the ``$2n+1$'' theorem, 
enables writing explicit formulas for first-order dispersion coefficients
that only need the \emph{uniform} field wavefunction response as an input.
Thus, one can take advantage of the already implemented linear-response 
tools to calculate a wide range of new materials properties, such as
flexoelectricity and the natural optical activity, at 
essentially no cost -- and without the need for explicitly implementing or 
calculating the wavefunction response to a gradient of the external field.
Finally, we demonstrate our formalism by implementing the 
formulas for the clamped-ion flexoelectric coefficients and the 
\emph{dynamical quadrupole tensor} (the higher-order multipolar counterpart 
of the Born dynamical charge tensor). 
The flexoelectric coefficients calculated for several materials are consistent with  
previously published results.~\cite{artcalc,Cyrus} The relationship,
established by R. M. Martin in his seminal paper,~\cite{Martin}
between the sublattice sum of the quadrupole moments and the clamped-ion 
piezoelectric coefficients is numerically verified to a high degree of
accuracy.
Both quantities converge with respect to plane-wave cutoff
and ${\bf k}$-mesh density comparably fast to ``standard'' linear-response 
properties (e.g. the macroscopic dielectric tensor), and can now be
obtained in a tiny fraction of the computational burden that was formerly needed.

This work is organized as follows. In Section~\ref{sec:theory} we present our 
method, based on the long-wavelength expansion of DFPT,
and provide general formulas for dispersion properties at the 
lowest orders in ${\bf q}$. 
In Section~\ref{sec:perturb} we discuss the 
finite-${\bf q}$ generalization of the electric-field response, 
which we shall use to define and compute the polarization response
in the long-wavelength limit.
In Sections~\ref{sec:quad} and~\ref{sec:flexo} we demonstrate our long-wave approach
by deriving and calculating the dynamical quadrupole and 
clamped-ion flexoelectric tensors in selected materials and model systems. 
Finally, in Section~\ref{sec:outlook} we present our conclusions and outlook, 
e.g. regarding future generalizations of our method to other dispersion 
properties.
The Appendices provide additional analytic support to the formulas reported
in the main text. 

\section{Long-wave perturbation theory}

\label{sec:theory}

\subsection{Density-functional perturbation theory}

Here we shall briefly introduce the basic principles of DFPT,
both for completeness and in order to support the formal
developments of the later sections. 
Consider an external perturbation to the electronic ground state, which we describe by assuming 
a parametric dependence of the Hamiltonian operator on a small parameter 
$\lambda$,
\begin{equation}
\hat{H}(\lambda) = \hat{H}^{(0)} + \lambda \hat{H}^{(1)} + \lambda^2 \hat{H}^{(2)} + \cdots .
\end{equation}
The linear response of the wavefunctions to the perturbation can be recast in terms of a
Sternheimer equation,
\begin{equation}
\hat{Q} \left( H^{(0)} - \epsilon^{(0)}_m \right) \hat{Q} | \psi^{(1)}_m \rangle = - \hat{Q} \hat{\mathcal{H}}^{(1)} | \psi^{(0)}_m \rangle,
\label{stern}
\end{equation}
where $\hat{Q}$ indicates the projector on the unoccupied band manifold, and
\begin{equation}
\hat{\mathcal{H}}^{(1)} = \hat{H}^{(1)} + \hat{V}^{(1)}
\end{equation}
contains, in addition to the external perturbation $ \hat{H}^{(1)}$, the 
self-consistent (SCF) potential response, $\hat{V}^{(1)}$, that depends on
the first-order electron density as
\begin{eqnarray}
V^{(1)}({\bf r}) &=& \int d^3 r' \, K_{\rm Hxc}({\bf r,r'}) n^{(1)}({\bf r}), \\
n^{(1)}({\bf r}) &=& 2 \Re \, \sum_m  \langle \psi^{(0)}_m | {\bf r} \rangle \langle {\bf r} |  \psi^{(1)}_m \rangle.
\end{eqnarray}
$K_{\rm Hxc}({\bf r,r'})$ is the Hartree, exchange and correlation (Hxc) kernel, which is defined as
the variation of the SCF potential at ${\bf r}$ with respect to a charge density perturbation
at ${\bf r}'$, calculated at the ground-state density ${n^{(0)}}$,
\begin{equation}
K_{\rm Hxc}({\bf r,r'}) =  \frac{\delta V_{\rm Hxc}({\bf r})}{\delta n({\bf r}')} \Big|_{n^{(0)}} = 
 \frac{\delta^2 E_{\rm Hxc}}{\delta n({\bf r}) \delta n({\bf r}')} \Big|_{n^{(0)}}.
\end{equation}
The second-order variation of the energy with respect to the perturbation can be then written
as
\begin{equation}
E^{(2)} = \sum_m  \langle \psi^{(0)}_m | \hat{H}^{(1)}  | \psi^{(1)}_m \rangle +  \frac{1}{2} \frac {\partial^2 E }{\partial \lambda^2 },
\label{nonvar}
\end{equation}
where the second term on the right-hand side does not depend on the first-order wavefunctions,
\begin{equation}
\frac{1}{2} \frac {\partial^2 E }{\partial \lambda^2 } = \sum_m \langle \psi^{(0)}_m | \hat{H}^{(2)}  | \psi^{(0)}_m \rangle.
\end{equation}

One can also recast the linear-response problem as a variational
functional of the first-order wavefunctions,~\cite{Gonze-95a}
\begin{eqnarray}
E^{(2)} &=& \sum_m \langle \psi^{(1)}_m | \left( H^{(0)} - \epsilon^{(0)}_m \right) | \psi^{(1)}_m \rangle \nonumber \\
        && + \sum_m \left( \langle \psi^{(1)}_m | H^{(1)}  | \psi^{(0)}_m \rangle + \langle \psi^{(0)}_m | H^{(1)}  | \psi^{(1)}_m \rangle \right) \nonumber \\
        && + \frac{1}{2} \int_\Omega \int K_{\rm Hxc}({\bf r,r'})  
           n^{(1)}({\bf r}) n^{(1)}({\bf r}')  d^3 r d^3 r' \nonumber \\
   && + \frac{1}{2} \frac {\partial^2 E }{\partial \lambda^2 }, 
\label{variational}
\end{eqnarray}   
(the double integral of the third line must be taken once over all space, and once 
over the primitive unit cell, whose volume is $\Omega$)     
to be solved within the ``parallel-transport gauge'' (i.e. under 
the constraint of orthonormality to the valence manifold, $\mathcal{V}$),
\begin{equation}
 \langle \psi^{(1)}_{j} | \psi^{(0)}_{l} \rangle = 0, \qquad j,l \in \mathcal{V}.
\end{equation}
Eq.~(\ref{nonvar}) and Eq.~(\ref{variational}) manifestly coincide if the first-order wavefunctions satisfy the Sternheimer equation,
Eq.~(\ref{stern}); however, the latter expression has the virtue of being stationary with respect to variations of $| \psi^{(1)}_m \rangle$,
and such a characteristic will have a key importance in the context of this work, as we shall see shortly.

\subsection{Unconstrained variational formulation}

First, recall the definition of the valence- and conduction-band projectors (we have already seen the 
latter in the previous subsection),
\begin{equation}
\hat{P} =  \sum_n | \psi^{(0)}_n \rangle \langle \psi^{(0)}_n|, \qquad \hat{Q} = 1-\hat{P}.
\end{equation}
We shall now use these definitions to write the linear response problem as 
an \emph{unconstrained} variational minimum of the following functional
\begin{eqnarray}
E^{(2)} &=& \sum_m \langle \psi^{(1)}_m | \left( \hat{H}^{(0)} + a \hat{P} - \epsilon^{(0)}_m \right) | \psi^{(1)}_m \rangle \nonumber \\
        && + \sum_m \langle \psi^{(1)}_m | \hat{Q} \, \hat{H}^{(1)}  | \psi^{(0)}_m \rangle + c.c.
        \nonumber \\
        && + \frac{1}{2} \int_\Omega \int K_{\rm Hxc}({\bf r,r'})  
           n^{(1)}({\bf r}) n^{(1)}({\bf r}')  d^3 r d^3 r' \nonumber \\
   && + \frac{1}{2} \frac {\partial^2 E }{\partial \lambda^2 },
\label{unconstrained}
\end{eqnarray}        
Note the explicit introduction of the band projectors in the first and second line,
and implicitly in the third line via a redefinition of the first-order electron density,
\begin{equation}
n^{(1)}({\bf r}) = \sum_m \langle \psi^{(1)}_m | \hat{Q} | {\bf r} \rangle \langle {\bf r} | \psi^{(0)}_m \rangle + c.c. .
\end{equation}

The parameter $a$ is a constant with the dimension of an energy, whose role is 
to ensure that the matrix element in the first line of Eq.~(\ref{unconstrained}), 
quadratic in the first-order wavefunctions, is defined positive, and hence that 
the functional is \emph{stable}. 
To see this, consider the mean value of the  
operator in the round brackets on a valence ($v$) or conduction ($c$) state,
\begin{eqnarray}
\langle \psi^{(0)}_{v} | 
                      (\hat H^{(0)} + a \hat{P} - \epsilon_{n} ) | \psi^{(0)}_{v} \rangle &=&
                 \epsilon_{v } + a - \epsilon_{n}, \\
\langle \psi^{(0)}_{c } | 
                      (\hat H^{(0)} + a \hat{P} - \epsilon_{n} ) | \psi^{(0)}_{c } \rangle &=&
                 \epsilon_{c } - \epsilon_{n }.
\end{eqnarray}
As $n$ belongs to the valence band, the matrix element on the conduction state is 
always positive independent of $a$. Regarding the valence state, for the
value $\epsilon_{v} + a - \epsilon_{n}$ to be guaranteed to be
positive it suffices to set $a$ to any positive energy that is larger than the total valence
bandwidth.

The insertion of a conduction band projector, $\hat{Q}$, 
in both the charge density and in the second line of Eq.~(\ref{unconstrained})
serves to enforcing the parallel-transport gauge, i.e. that 
at the variational minimum the solutions $\psi^{(1)}$ be strictly orthogonal 
to the valence manifold. 
It is easy to see how this works: Thanks to the projectors $\hat{Q}$, the addition of a 
small valence component to the trial solution $\psi^{(1)}$ leaves the energy unaltered 
\emph{except} for the (quadratic) matrix element in the first line of Eq.~(\ref{unconstrained}).
The latter, in turn, always provides a positive contribution to the energy, whose 
magnitude depends on the parameter $a$. 
Therefore, $a$ has no influence other than preventing the first-order wavefunctions from acquiring 
arbitrarily large components on the valence manifold, which would lead to runaway solutions.

Following these considerations, it is not difficult to get convinced that the variational solution 
of this unconstrained energy functional is unique, and corresponds precisely  to the constrained 
minimization procedure described by Gonze~\cite{Gonze}. 
It also leads, by differentiating Eq.~(\ref{unconstrained}) with respect to $\langle \psi^{(1)}_m |$, to the form of the 
Sternheimer equation proposed by Baroni and coworkers~\cite{Baroni-01},
\begin{equation}
\left( \hat{H}^{(0)} + a \hat{P} - \epsilon^{(0)}_m \right) | \psi^{(1)}_m \rangle = -\hat{Q} \, \hat{\mathcal{H}}^{(1)}  | \psi^{(0)}_m \rangle.
\end{equation}
Such a form clearly enforces $\hat{P} | \psi^{(1)}_m \rangle = 0$, and reduces 
to Eq.~(\ref{stern}) once the left-hand side is projected on the conduction manifold.

\subsection{Factorization of the phase}

To appreciate the practical advantages of the unconstrained formulation 
of the previous subsection, we shall now apply it to a monochromatic 
perturbation in a periodic crystal. This can be expressed as a phase
times a cell-periodic part,
\begin{equation}
\hat{H}^{(1)}({\bf r,r}') = e^{i{\bf q\cdot r}} \hat{H}^{(1)}_{\bf q}({\bf r,r}')
\end{equation}
As customary, we shall work with the cell-periodic part of the Bloch wavefunctions by writing
\begin{displaymath}
\psi_{m \bf k}({\bf r}) = e^{i{\bf k\cdot r}} u_{m \bf k}({\bf r}),
\end{displaymath}
which allows one to reabsorb the incommensurate phase $e^{i{\bf q\cdot r}}$ 
by performing appropriate shifts of the states and operators in momentum space.

For the sake of generality, we shall consider the mixed derivative
with respect to two distinct perturbations, $\lambda_1$ and $\lambda_2$, whose physical
nature will be specified later in this manuscript.
(The functional, strictly speaking, is variational only for $\lambda_1=\lambda_2$; 
yet, even in the mixed case it preserves the stationary character with respect to small
variations in the first-order wavefunctions.)
We shall implicitly assume that the crystal under study is a time-reversal (TR) symmetric
insulator. (A generalization of the formulas to TR-broken materials, while not difficult,
would have unnecessarily complicated the notation.)
The second-order energy can be written then as
\begin{eqnarray}
E^{\lambda_1^* \lambda_2}_{\bf q}  &=& s \int_{\rm BZ} [d^3 k] \, \sum_m  E^{\lambda_1^* \lambda_2}_{m\bf k,q} \nonumber \\
   && + \frac{1}{2} \int_\Omega \int K_{\bf q}({\bf r,r'})  
           n_{\bf q}^{\lambda_1 *}({\bf r}) n_{\bf q}^{\lambda_2}({\bf r}')  d^3 r d^3 r' \nonumber \\
   && + \frac{1}{2} \frac {\partial^2 E }{\partial \lambda_1^*  \partial \lambda_2},
\label{mixedq}
\end{eqnarray}
where the quantity in the first line is given by 
\begin{eqnarray}
E^{\lambda_1^* \lambda_2}_{m\bf k,q} &=& \langle u^{\lambda_1}_{m {\bf k,q}} | 
        \left( \hat H^{(0)}_{\bf k+q} + a \hat{P}_{\bf k+q} - \epsilon_{m {\bf k}} \right) | u^{\lambda_2}_{m {\bf k,q}} \rangle  \nonumber \\
    && + \langle u^{\lambda_1}_{m {\bf k,q}} | \hat{Q}_{\bf k+q} 
     \hat H^{\lambda_2}_{\bf k, q}  | u^{(0)}_{m {\bf k}} \rangle  \nonumber \\
     && +  \langle  u^{(0)}_{m {\bf k}}| \left( \hat H^{\lambda_1 }_{\bf k, q} \right)^\dagger  \hat{Q}_{\bf k+q} 
      | u^{\lambda_2}_{m {\bf k,q}} \rangle,
\label{mixedq_mk}
\end{eqnarray}  
$s=2$ is the spin multiplicity and we have used the following shorthand notation for the
Brillouin-zone averages,
\begin{displaymath}
\int_{\rm BZ} [d^3 k] = \frac{\Omega}{(2\pi)^3} \int_{\rm BZ} d^3 k.
\end{displaymath} 
The last (third) line in Eq.~(\ref{mixedq}) is, as usual, the nonvariational contribution to the 
second-order energy, while the second line
contains the self-consistent energy that depends quadratically on the first-order electron densities~\footnote{
  The following expression is only justified in time-reversal (TR) symmetric materials.
  For simplicity, we shall assume TR symmetry throughout this work.}
\begin{equation}
n^{\lambda}_{\bf q}({\bf r}) = 2s \int_{\rm BZ} [d^3 k] \, \sum_m
       \langle  u^{(0)}_{m {\bf k}} | {\bf r} \rangle \langle {\bf r} | \hat{Q}_{\bf k+q} | u^{\lambda}_{m {\bf k,q}} \rangle.
\end{equation}
Note that we have introduced new symbols for the phase-corrected Hxc kernel (we shall 
specialize to the local density approximation, LDA),
\begin{displaymath}
K_{\bf q}({\bf r,r'}) = K_{\rm Hxc}({\bf r,r'}) e^{i{\bf q} \cdot ({\bf r'-r})},
\end{displaymath}
the operators in momentum space,
\begin{displaymath}
\hat{O}_{\bf k} = e^{-i{\bf k} \cdot {\bf r}} \hat{O} e^{i{\bf k} \cdot {\bf r}'},
\end{displaymath}
and the cell-periodic part of the charge-density response
\begin{displaymath}
n^{\lambda}_{\bf q}({\bf r})= e^{-i{\bf q} \cdot {\bf r}} n^{\lambda}({\bf r}).
\end{displaymath}

From these formulas, one can now appreciate the most remarkable property of the 
unconstrained functional: Unlike the original version, where the
orthonormality constraint is taken by calculating the scalar 
products with ground-state valence orbitals at ${\bf k+q}$,
the present version is written in a manifestly gauge-invariant form, i.e. only 
\emph{operators} explicitly depend on ${\bf q}$.
This is a key advantage when developing a perturbative theory in 
${\bf q}$, as the derivatives of the operators in momentum space
are well-defined mathematical objects, and do not suffer from the 
phase indeterminacy of the Bloch states.

\subsection{``$2n+1$'' theorem \label{sec_2np1}}

At this point, we can treat $E^{\lambda_1^* \lambda_2}_{\bf q}$ as a new functional of
$| u^{\lambda_{1,2}}_{m {\bf k,q}} \rangle$, which depends parametrically on ${\bf q}$.
We can then take advantage of the established mathematical tools of perturbation theory 
to expand $E^{\lambda_1^* \lambda_2}_{\bf q}$ in powers of ${\bf q}$ around ${\bf q}=0$,
which has the physical interpretation of a long-wave expansion.
This can be pushed, in principle, to any order in ${\bf q}$. In particular, in virtue
of the ``$2n+1$'' theorem,~\cite{Gonze-95} the knowledge of the ${\bf q}$-derivatives of the wavefunctions
up to order $n$ is sufficient to calculate response properties up to $\mathcal{O}(q^{2n+1})$.
As we shall see in the following, this is especially useful at the lowest orders: The
computational tools to calculate the $n=0$ (and, sometimes, $n=1$) response functions are 
already available in many public first-principles packages, which implies that many response 
properties can be, in principle, extracted
\emph{without even implementing a new response function} in the code.
(In the following, we shall illustrate this strategy at a formal level, without
specifying the physical nature of the perturbations; practical examples will be
provided in Sections~\ref{sec:quad} and~\ref{sec:flexo}.)

At first order in ${\bf q}$ the ``$2n+1$'' theorem reduces to the Hellmann-Feynman
theorem and can be summarized as follows,
\begin{equation}
E^{\lambda_1^* \lambda_2}_{\gamma} = \frac{d E^{\lambda_1^* \lambda_2}_{\bf q}}{d q_\gamma} \Big|_{{\bf q}=0}= 
                               \frac{\partial E^{\lambda_1^* \lambda_2}_{\bf q}}{\partial q_\gamma} \Big|_{{\bf q}=0},
     \label{grade}                          
\end{equation}
which states that the ${\bf q}$-gradients of the response functions $| u^{\lambda}_{m {\bf k,q}} \rangle$ 
are not needed to access the ${\bf q}$-gradient of the stationary second-order functional.
(We specialize our formulas to a neighborhood of ${\bf q}=0$, as such a limit is 
directly relevant for the macroscopic response properties of the crystal.) 
In particular, we have
\begin{eqnarray}
E^{\lambda_1^* \lambda_2}_{\gamma} &=& s \int_{\rm BZ} [d^3 k] \, \sum_m  E^{\lambda_1^* \lambda_2}_{m\bf k,\gamma} \nonumber \\
         && + \frac{1}{2} \int_\Omega \int K_{\gamma}({\bf r,r'})  
           n^{\lambda_1 *}({\bf r}) n^{\lambda_2}({\bf r}')  d^3 r d^3 r' \nonumber \\
   && + \frac{1}{2} \frac{\partial }{\partial q_\gamma} \left( \frac {\partial^2 E }{\partial \lambda_1^*  \partial \lambda_2} \right) \Big|_{{\bf q}=0},
\label{e2gamma}
\end{eqnarray}
where we have used a short-hand notation for the ${\bf q}$-derivative of the 
Hartree and exchange-correlation kernel, 
\begin{equation}
K_\gamma({\bf r,r'}) =\frac{ \partial K_{\bf q}({\bf r,r'})}{\partial q_\gamma} \Big|_{{\bf q}=0},
\end{equation}
and the band-resolved contribution reads as
\begin{equation}
\begin{split}
 & E^{\lambda_1^* \lambda_2}_{m\bf k,\gamma} =
   \langle u^{\lambda_1}_{m {\bf k}} | \partial_\gamma \hat H^{(0)}_{\bf k} | u^{\lambda_2}_{m {\bf k}} \rangle  \\ 
 & \quad + \langle u^{\lambda_1}_{m {\bf k}} | \partial_\gamma \hat{Q}_{\bf k} \, \hat{\mathcal{H}}^{\lambda_2}_{\bf k}  | u^{(0)}_{m {\bf k}} \rangle  
  +  \langle  u^{(0)}_{m {\bf k}}| \left( \hat{\mathcal{H}}^{\lambda_1 }_{\bf k} \right)^\dagger 
              \partial_\gamma \hat{Q}_{\bf k} | u^{\lambda_2}_{m {\bf k}} \rangle \\
  & \quad + \langle u^{\lambda_1}_{m {\bf k}} |  \hat{H}^{\lambda_2}_{\bf k,\gamma}  | u^{(0)}_{m {\bf k}} \rangle  
  +  \langle  u^{(0)}_{m {\bf k}}| \left( \hat{H}^{\lambda_1 }_{\bf k,\gamma} \right)^\dagger  | u^{\lambda_2}_{m {\bf k}} \rangle.
\end{split}
\label{e2gamma_mk}
\end{equation}  
Here we have introduced new symbols for the ${\bf q}$-derivatives of the external perturbation,
\begin{equation}
\hat{H}^{\lambda }_{\bf k,\gamma} = \frac{ \partial \hat{H}^{\lambda }_{\bf k,q}}{\partial q_\gamma} \Big|_{{\bf q}=0},
\end{equation}
and for the ${\bf k}$-derivatives of the ground-state operators (Hamiltonian or band projectors), e.g.,
\begin{equation}
\partial_\gamma \hat H^{(0)}_{\bf k} = \frac{ \partial \hat{H}^{(0)}_{\bf k+q}}{\partial q_\gamma} \Big|_{{\bf q}=0}.
\end{equation}
Also, we have removed the ${\bf q}$ subscript from those quantities (either 
first-order wave functions, densities or perturbing operators) that are intended to be calculated at 
${\bf q}=0$, e.g.,
\begin{equation}
| u^{\lambda}_{m {\bf k}} \rangle = | u^{\lambda}_{m {\bf k,q}=0} \rangle.
\end{equation}

Note that we have used the symbol $\hat{\mathcal{H}}$ in the second line of Eq.~(\ref{e2gamma_mk})
to indicate that the self-consistent (SCF) Hartree and exchange-correlation potential must be included 
in the first-order Hamiltonian at ${\bf q}=0$.
The SCF part of $\hat{\mathcal{H}}^{\lambda}_{\bf k}$ comes from the Hartree and exchange-correlation
term in Eq.~(\ref{mixedq}) via the partial derivative of the first-order density with respect to
$q_\gamma$,
\begin{equation}
\frac{\partial n^{\lambda}({\bf r})}{\partial q_\gamma} =  2s \int_{\rm BZ} [d^3 k] \, \sum_m
       \langle  u^{(0)}_{m {\bf k}} | {\bf r} \rangle \langle {\bf r} | \partial_\gamma \hat{Q}_{\bf k} | u^{\lambda}_{m {\bf k}} \rangle.
\label{partialrho}
\end{equation}
Crucially, the SCF potential needs to be explicitly calculated only at the level of the $\mathcal{O}(q^0)$
perturbation; the ${\bf q}$-gradient of the perturbation, in the third line, only concerns the \emph{external
potential} part, $\hat{H}$.
(This is, again, a consequence of the ``$2n+1$'' theorem.)
Note also that $\partial_\gamma \hat{P}_{\bf k}$ only has cross-gap matrix elements,
thus it doesn't contribute to the first line of Eq.~(\ref{e2gamma_mk}), and 
that we could omit $\hat Q_{\bf k}$ from the matrix elements in the third line, as it always
appeared next to a conduction-band state.

The above formulas enable the calculation of the ``$d/dq_\gamma$'' response with
computational workload that is comparable to the uniform (${\bf q}=0$) case.
Indeed, 
only the ${\bf q}=0$ first-order wavefunctions are needed as ingredients;
the additional burden consists in the implementation of the new operators that
appear in Eqs.~(\ref{e2gamma}) and~(\ref{e2gamma_mk}), but once this is done the
evaluation of the corresponding matrix elements proceeds at essentially no cost.
Most of these ``new'' operators are, in fact, well known in the context of band theory,
and are standard in most DFPT implementations (e.g. the velocity operator, 
$\partial_\gamma \hat H^{(0)}_{\bf k}$, or the derivatives of the band projectors).
For example, the second line of Eq.~(\ref{e2gamma_mk}) might look unusual at
first sight, but it can be made more explicit by observing that
$\partial_\gamma \hat{Q}_{\bf k} = -\partial_\gamma \hat{P}_{\bf k}$, and that 
\begin{equation}
\partial_\gamma \hat{P}_{\bf k} = \sum_n \left(  |u^{(0)}_{n {\bf k}} \rangle \langle \tilde{\partial}_\gamma u^{(0)}_{n {\bf k}} |  +
                                           | \tilde{\partial}_\gamma u^{(0)}_{n {\bf k}} \rangle  \langle  u^{(0)}_{n {\bf k}} |  \right),
\end{equation}
where $| \tilde{\partial}_\gamma u^{(0)}_{n {\bf k}} \rangle$ are the ``covariant derivatives'' of
the ground-state wavefunctions (also known as ``$d/dk_\gamma$'' response functions), and are
orthogonal to the valence manifold.
Then one immediately obtains
\begin{equation}
\begin{split}
 & \langle u^{\lambda_1}_{m {\bf k}} | \partial_\gamma \hat{Q}_{\bf k} \, \hat{\mathcal{H}}^{\lambda_2}_{\bf k}  | u^{(0)}_{m {\bf k}} \rangle = \\ 
 &\quad -\sum_n \langle u^{\lambda_1}_{m {\bf k}} |  \tilde{\partial}_\gamma u^{(0)}_{n {\bf k}} \rangle  
 \langle  u^{(0)}_{n {\bf k}} |  \hat{\mathcal{H}}^{\lambda_2}_{\bf k}  | u^{(0)}_{m {\bf k}} \rangle,
 \end{split}
\end{equation}
which is now a rather familiar expression in the context of DFPT.

The truly new pieces in Eqs.~(\ref{e2gamma}) and~(\ref{e2gamma_mk}) are the ${\bf q}$-derivatives 
of the monochromatic perturbation, and the ${\bf q}$-derivative of the SCF kernel. 
We shall defer 
the discussion of the former, which depends on the specific perturbation, to Sections~\ref{sec:quad} and~\ref{sec:flexo}.
The latter is particularly simple to evaluate in
the framework of the local-density approximation (LDA),
where the XC part does not contribute (it is 
independent of ${\bf q}$).
As we are only left with electrostatic effects, it is most convenient to work
in reciprocal space, where the Coulomb (Hartree) kernel is local,
\begin{equation}
K_{\rm H,\bf q}({\bf G,G'}) = 4\pi \frac{\delta_{\bf GG'}}{|{\bf G+q}|^2}.
\end{equation}
(${\bf G}$ and ${\bf G}'$ stand for reciprocal-lattice vectors, and $\delta$ is
a Kronecker symbol.) The ${\bf q}$-gradient (at ${\bf q}=0$) 
of the above expression is easily computed,
\begin{equation}
K_\gamma({\bf G,G'}) = -8\pi G_\gamma \frac{\delta_{\bf GG'}}{G^4}.
\end{equation}
The ${\bf G}=0$ term must be, of course, excluded; this corresponds to
adopting short-circuit electrical boundary conditions, which is the correct 
choice for computing materials properties that have a tensorial nature.
(A formal justification of this point was provided 
in Ref.~\onlinecite{Gonze} for the uniform electric field problem, and in 
Ref.~\onlinecite{artlin} for the flexoelectric tensor.)

\subsection{Higher orders}

As we said, the ``$2n+1$'' theorem, in principle, provides access to 
the long-wave expansion terms of a given crystal response to any
order in ${\bf q}$. 
In general, the analytic formulas for higher orders in ${\bf q}$
can become rather cumbersome to derive, as they involve a larger number
of terms; plus, they typically require additional response functions to be 
implemented and calculated.
There is, however, an important exception to this statement that is worth 
discussing, as it is central to the topics that will be presented in the later 
sections.
Indeed, there are some notable cases where a perturbation produces a vanishing response 
at ${\bf q}=0$, and the interesting physics occurs only \emph{at first order} in 
${\bf q}$.
A classic example is that of a scalar potential perturbation: At ${\bf q}=0$,
the perturbation is a rigid shift of the potential reference, which has obviously 
no effect on the electronic structure; at first order in ${\bf q}$, one obtains
the response to a uniform electric field.~\cite{Gonze}
In such cases, the formula for the $\mathcal{O}(q^2)$ response 
simplifies considerably and, in fact, is only marginally more 
complicated than the first-order formulas, Eqs.~(\ref{e2gamma}) and~(\ref{e2gamma_mk}).

To be more specific, consider the following mixed derivative,
\begin{equation}
E^{\lambda_1^* \lambda_2}_{\gamma \delta} = \frac{d^2 E^{\lambda_1^* \lambda_2}_{\bf q}}{d q_\gamma dq_\delta} \Big|_{{\bf q}=0},
\end{equation}
and assume that 
\begin{equation}
| u^{\lambda_2}_{m {\bf k,q}=0} \rangle = 0.
\label{q0vanish}
\end{equation}
Consistent with the above notation, we shall indicate the response function to a gradient 
of the perturbation $\lambda_2$ as
\begin{equation}
| u^{\lambda_2}_{m {\bf k},\delta} \rangle = \left| \frac{\partial u^{\lambda_2}_{m {\bf k,q}}}{\partial q_\delta}\bigg|_{{\bf q}=0 } \right\rangle.
\end{equation}
Then we have 
\begin{equation}
E^{\lambda_1^* \lambda_2}_{\gamma \delta} =  
            \widetilde{E}^{\lambda_1^* \lambda_2}_{\gamma \delta}   +   
            \widetilde{E}^{\lambda_1^* \lambda_2}_{\delta \gamma}, 
\end{equation}
where the tilded (unsymmetrized) quantities read as
\begin{eqnarray}
\widetilde{E}^{\lambda_1^* \lambda_2}_{\gamma \delta} &=& 
      s \int_{\rm BZ} [d^3 k] \, \sum_m  \widetilde{E}^{\lambda_1^* \lambda_2}_{m\bf k,\gamma\delta} \nonumber \\
         && + \frac{1}{2} \int \int K_{\gamma}({\bf r,r'})  
           n^{\lambda_1 *}({\bf r}) n^{\lambda_2}_\delta ({\bf r}')  d^3 r d^3 r' \nonumber \\
   && + \frac{1}{4} \frac{\partial^2 }{\partial q_\gamma \partial q_\delta} 
           \left( \frac {\partial^2 E }{\partial \lambda_1^*  \partial \lambda_2} \right) \Big|_{{\bf q}=0},
\label{e2gamdel}
\end{eqnarray}
with
\begin{equation}
\begin{split}
 & \widetilde{E}^{\lambda_1^* \lambda_2}_{m\bf k,\gamma\delta} =
   \langle u^{\lambda_1}_{m {\bf k}} | \partial_\gamma \hat H^{(0)}_{\bf k} | u^{\lambda_2}_{m {\bf k},\delta} \rangle  \\ 
 & \quad + \frac{1}{2} \langle u^{\lambda_1}_{m {\bf k}} | \partial_{\gamma \delta} \hat{Q}_{\bf k} \, 
                                  \hat{\mathcal{H}}^{\lambda_2}_{\bf k}  | u^{(0)}_{m {\bf k}} \rangle \\
 & \quad + \langle u^{\lambda_1}_{m {\bf k}} | \partial_\gamma \hat{Q}_{\bf k} \, \hat{\mathcal{H}}^{\lambda_2}_{\bf k,\delta}  | u^{(0)}_{m {\bf k}} \rangle  
  +  \langle  u^{(0)}_{m {\bf k}}| \left( \hat{\mathcal{H}}^{\lambda_1 }_{\bf k} \right)^\dagger 
              \partial_\gamma \hat{Q}_{\bf k} | u^{\lambda_2}_{m {\bf k},\delta} \rangle \\
  & \quad + \frac{1}{2} \langle u^{\lambda_1}_{m {\bf k}} | \hat{H}^{\lambda_2}_{\bf k,\gamma\delta}  | u^{(0)}_{m {\bf k}} \rangle  
  +  \langle  u^{(0)}_{m {\bf k}}| \left( \hat{H}^{\lambda_1 }_{\bf k,\gamma} \right)^\dagger  | u^{\lambda_2}_{m {\bf k},\delta} \rangle,
\end{split}
\label{e2gamdel_mk}
\end{equation}  
and
\begin{equation}
n^{\lambda_2}_{\delta}({\bf r}) = 2s \int_{\rm BZ} [d^3 k] \, \sum_m
       \langle  u^{(0)}_{m {\bf k}} | {\bf r} \rangle \langle {\bf r} | u^{\lambda_2}_{m {\bf k,\delta}} \rangle.
\end{equation}
(Again, we could drop the conduction-band projector 
as $| u^{\lambda_2}_{m {\bf k,\delta}} \rangle$ belongs to the conduction band 
by construction.)
The resulting formulas for the second-order energy are essentially identical to those
derived in Section~\ref{sec_2np1} for the first order in ${\bf q}$, with 
three main differences: 
(i) the result needs now to be symmetrized with respect to
$\gamma$ and $\delta$; (ii) every occurrence of the response functions and
perturbing operators that depend on $\lambda_2$ need to be replaced with their 
next higher-order gradient in ${\bf q}$; (iii) there is a new term in 
Eq.~(\ref{e2gamdel_mk}) containing the \emph{second} ${\bf k}$-gradient
of the band projector, $\partial_{\gamma \delta} \hat{Q}_{\bf k}$. The
latter is multiplied by $\hat{\mathcal{H}}^{\lambda_2}_{\bf k}$, which 
we have included to account for cases where the perturbation $\lambda_2$,
while yielding a vanishing response at ${\bf q}=0$, may not vanish therein.

Similar considerations can be used in order to push the expansion to $\mathcal{O}(q^3)$
whenever \emph{both} perturbations, $\lambda_1$ and $\lambda_2$, satisfy Eq.~(\ref{q0vanish}).

\section{Treatment of the polarization response}

\label{sec:perturb}

Many materials properties (including the flexoelectric tensor, discussed
in Section~\ref{sec:flexo}) involve, in one way or the other, the polarization 
response to an external perturbation.
Correctly treating the long-wavelength limit of the electrical polarization 
is far from trivial in the framework of density-functional perturbation theory.
In presence of a spatial modulation the standard formulas (e.g., based on the 
Berry-phase approach) are not applicable, since the latter are specialized to the
macroscopic response at the Brillouin zone center.
To work around this issue, in Ref.~\onlinecite{Cyrus} the polarization response 
to some monochromatic external field, $\lambda^{\bf q}$, was expressed as the 
current-density (${\bf J}$) response to the time derivative of the field, 
\begin{equation}
\frac{d {\bf P}^{\bf q}}{d \lambda^{\bf q}} = \frac{d {\bf J}^{\bf q}}{d  \dot{\lambda}^{\bf q}}.
\end{equation}
In a quantum-mechanical context, this can be expressed~\cite{Cyrus,metric} via the following formula,
\begin{equation}
\frac{d P_\alpha^{\bf q}}{d \lambda^{\bf q}} = 
  \frac{2 s}{\Omega} \int_{\rm BZ} [d^3 k] \, \sum_m  \langle u^{(0)}_{m\bf k} | \hat{J}_{\alpha \bf k, q} | \delta u^\lambda_{m\bf k, q} \rangle,
\label{adiab}
\end{equation}
where $\hat{J}_{\alpha \bf k, q}$ is the current-density operator at a given value of
${\bf q}$, $| \delta u^\lambda_{m\bf k, q} \rangle$ describe the \emph{adiabatic} wavefunction
response~\cite{Cyrus} to the perturbation velocity in the limit of $\dot{\lambda}^{\bf q} \rightarrow 0$
and the index $m$ runs over the occupied manifold.
(In other words, if we modulate the perturbation in time with a dynamical phase
$e^{-i\omega t}$, $| \delta \psi^{\bf q}_i \rangle$ is related to 
the first-order term in the low-frequency expansion of the wavefunction 
response.)

Unfortunately, Eq.~(\ref{adiab}) is not directly useful to our scopes, as it is 
not explicitly written as a second derivative of the total energy.
To circumvent this issue, we shall use the known relationship 
between the electric field and the polarization,~\cite{fixedd} 
${\bf P}= -\partial E / \partial \bm{\mathcal{E}}$, to rewrite
${\bf P}$ as a mixed derivative with respect to 
the electric field and the external perturbation $\lambda$,
\begin{equation}
\frac{d P_\alpha^{\bf q}}{d \lambda^{\bf q}} =  -\frac{d^2 E}{d \mathcal{E}^{\bf -q}_\alpha d \lambda^{\bf q}}.
\label{pfrome}
\end{equation}
This strategy recovers the established DFPT formulas~\cite{Gonze/Lee} for 
the polarization response in the ${\bf q}=0$ case. [For instance,
if $\lambda$ is an atomic displacement, Eq.(\ref{pfrome}) reduces
to the standard linear-response expression for the Born effective charge tensor.]
It presents, however, a new complication in that we need to generalize the 
electric-field perturbation to finite values of ${\bf q}$.
To do that, we shall express the $\bm{\mathcal{E}}$-field perturbation
as the time derivative of the $\bf A$-field perturbation, again by means
of adiabatic perturbation theory.
As we shall see shortly, this will allow us to write the polarization 
response in a variational form, and therefore apply the formalism
developed in the previous sections to perform its long-wave expansion.

In the following subsections we shall first discuss the response to a 
monochromatic \emph{vector potential} at finite ${\bf q}$, which is the 
fundamental building block of our approach.
The \emph{electric field} response is then defined as the frequency 
derivative of the vector potential response via a first-order expansion
in the frequency.
Finally, we shall discuss the simpler case of the \emph{scalar potential}
perturbation and show that, at first order in ${\bf q}$, it correctly recovers the 
electric-field response (as defined via adiabatic perturbation theory) at ${\bf q}=0$.

\subsection{Vector potential}

The problem will be broken down into into two separate steps: First, we shall 
review the coupling of a generic Hamiltonian to an external ${\bf A}$-field
in the linear regime, following the guidelines of Ref.~\onlinecite{Cyrus}. 
Next, we shall discuss the response to such
a perturbation with special attention to the lowest orders in ${\bf q}$.
In particular we shall see that, at first order in ${\bf q}$, 
the wave function response can be written in terms of second ${\bf k}$-gradients of
the ground-state orbitals, plus the orbital response to a uniform magnetic field.

\subsubsection{Coupling to a vector potential field}

The coupling of a generic Hamiltonian to a vector potential field can be written 
as~\cite{Mauri/Louie:96,Essin},
\begin{equation}
\hat{H}({\bf r}, {\bf r}') = \hat{H}^{(0)}({\bf r}, {\bf r}') e^{i \mathcal{Q} \int_{\bf r'}^{\bf r} {\bf A} \cdot d \bm{\ell}},
\label{coupling}
\end{equation}
where the line integral is assumed to be taken along the straight 
path connecting ${\bf r}'$ to ${\bf r}$, $\mathcal{Q}$ is the particle 
charge ($\mathcal{Q}=-e$ for electrons), and we use atomic units, i.e. we 
have set $\hbar=c=1$.
The linear expansion of the above expression in powers of the vector 
potential components to first order yields
\begin{equation}
\hat{H}({\bf r}, {\bf r}') \simeq \hat{H}^{(0)}({\bf r}, {\bf r}') 
  + i \mathcal{Q} \hat{H}^{(0)}({\bf r}, {\bf r}') \int_{\bf r'}^{\bf r} {\bf A} \cdot d \bm{\ell}.
\end{equation}
We shall consider a monochromatic ${\bf A}$-field, written as a real constant
times a complex phase of wavevector ${\bf q}$,
\begin{equation}
A_\alpha ({\bf r}) =  \lambda_\alpha e^{i {\bf q \cdot r}}.
\end{equation}
After taking the line integral, one obtains a closed expression for the first-order 
Hamiltonian in a coordinate representation,
\begin{equation}
\hat{H}^{A_\alpha}_{\bf q} ({\bf r}, {\bf r}') = i \mathcal{Q} \hat{H}^{(0)} ({\bf r}, {\bf r}') (r_\alpha - r'_\alpha) \, 
  \frac{  e^{i {\bf q \cdot r}} - e^{i {\bf q \cdot r'}} } {i {\bf q \cdot ({\bf r - r'})} }.
\end{equation}
Note that the first-order Hamiltonian is related to the current-density operator as
\begin{equation}
\hat{J}_\alpha({\bf -q})  = -\hat{H}^{A_\alpha}_{\bf q},
\label{jalp}
\end{equation}
which stems from the thermodynamic relationship ${\bf J}({\bf r}) = -\delta E / \delta {\bf A}({\bf r})$.

In the long-wavevelength context of this work, it is useful to expand the fraction 
on the right-hand side in powers of ${\bf q}$, and to move the incommensurate phase 
factor to the left,
\begin{equation}
 \frac{  e^{i {\bf q \cdot r}} - e^{i {\bf q \cdot r'}} } {i {\bf q \cdot ({\bf r - r'})} } 
  = e^{i {\bf q \cdot r}} \sum_{n=0}^\infty \frac{(-i)^n}{(n+1)!}  \, [{\bf q \cdot (r-r')}]^n.
\end{equation}
The above expansion clarifies that the fraction simplifies to unity in the ${\bf q}=0$ limit,
where the first-order Hamiltonian reads as
\begin{equation}
\hat{H}^{A_\alpha}_{{\bf q}=0} = -i \mathcal{Q} \left[ \hat{H}^{(0)}, r_\alpha \right].
\end{equation}
This provides a first ``sanity check'' of the present formalism: In the ${\bf q}=0$ limit
the current-density operator as defined in Eq.~(\ref{jalp}) correctly reduces to the velocity 
operator times the particle charge,
\begin{equation}
\hat{J}_\alpha = \mathcal{Q} \hat{v}_\alpha, \qquad \hat{v}_\alpha = i \left[ \hat{H}^{(0)}, r_\alpha \right].
\end{equation}

To make further progress towards a practical formalism, it is useful to consider the 
cell-periodic part of the first-order Hamiltonian in momentum space,
\begin{equation}
\hat{H}^{A_\alpha}_{\bf k, q} ({\bf r}, {\bf r}') = e^{-i({\bf k+q}) \cdot {\bf r}} \, \hat{H}^{A_\alpha}_{\bf q}({\bf r}, {\bf r}') \, 
          e^{i{\bf k} \cdot {\bf r}'}.
\end{equation}
This is a self-adjoint operator at any ${\bf q}$, and 
can be conveniently written as 
\begin{equation}
\hat{H}^{A_\alpha}_{\bf k, q} = -\mathcal{Q} \sum_{n=0}^{\infty} 
\left[ \sum_{\beta_1,\ldots,\beta_n} \frac{q_{\beta_1} \ldots q_{\beta_n}}{(n+1)!} 
\hat{H}_{\alpha \beta_1 \ldots \beta_n}  \right],
\end{equation}
where the individual terms in the summation stem from the 
${\bf k}$-expansion of the unperturbed Hamiltonian,
\begin{equation}
\hat{H}_{\alpha \beta_1 \ldots \beta_n} = \frac{ \partial^{n+1} \hat{H}^{(0)}_{\bf k} } 
{ \partial k_\alpha \partial k_{\beta_1} \ldots \partial k_{\beta_n}}.
\end{equation}
It is instructive to consider the special case of a local
Hamiltonian, where all expansion terms vanish except the lowest 
two,
\begin{equation}
\hat{H}^{A_\alpha, \rm loc}_{\bf k, q}  = -\mathcal{Q} \left( \hat{p}_{\bf k \alpha} + \frac{q_\alpha}{2}  \right).
\end{equation}

\subsubsection{Linear response to ${\bf A}$}

At this point, one could write a variational functional of the 
first-order wavefunction response to a monochromatic ${\bf A}$-field,
and follow the strategies outlined earlier in this work to 
calculate, e.g. the magnetic susceptibility of the crystal.
As our main focus here is on the electric field response, 
however, we shall skip this topic and directly focus on the 
wavefunction response to an ${\bf A}$-field -- this is the
crucial ingredient for what follows.
The wavefunction response can be written in terms of the 
following Sternheimer equation ,
\begin{equation}
\begin{split}
 \left( \hat H^{(0)}_{\bf k+q} + a \hat{P}_{\bf k+q} - \epsilon_{m {\bf k}} \right)  | u^{A_\alpha}_{m {\bf k,q}} \rangle = 
  -\hat{Q}_{\bf k+q} & \hat{H}^{A_\alpha}_{\bf k, q} | u^{(0)}_{m {\bf k}} \rangle  ,
  \label{stern-A}
\end{split}
\end{equation} 
(Note the absence of the SCF potential contribution, as a static vector potential field
leaves the charge density of the crystal unaltered in the linear regime by time reversal
symmetry.)
In the context of this work, we shall only need the zero-th and first orders
in the ${\bf q}$-expansion of $| u^{A_\alpha}_{m {\bf k,q}} \rangle$. Regarding
the ${\bf q}=0$ limit, it is easy to show that (since we are dealing with electrons we 
shall assume $\mathcal{Q}=-1$ henceforth)
\begin{equation}
| u^{A_\alpha}_{m {\bf k,q}=0} \rangle = \partial_\alpha \hat{P}_{\bf k} | u^{(0)}_{m {\bf k}} \rangle = | \tilde{\partial}_\alpha u^{(0)}_{m {\bf k}} \rangle,
 \label{ua0}
\end{equation}
where the ``$\partial$'' sign is a shortcut for the gradient in ${\bf k}$-space, and the tilde
indicates covariant derivation in the language of band theory.
Regarding the first order in ${\bf q}$, we shall report here the final result (a detailed derivation is reported in the Appendix)
\begin{equation}
\begin{split}
& | u^{A_\beta}_{m {\bf k,\gamma}} \rangle =  \left| \frac{ \partial u^{A_\beta}_{m {\bf k,q}=0} }{\partial q_\gamma } \right\rangle = \\
& \quad \frac{1}{2} \left( \partial^2_{\beta \gamma} \hat{P}_{\bf k} | u^{(0)}_{n {\bf k}} \rangle 
 -\left[ \partial_\gamma \hat{P}_{\bf k} , \partial_\beta \hat{P}_{\bf k} \right] | u^{(0)}_{n {\bf k}} \rangle +  
 | u^{\rm CG}_{n {\bf k},\beta \gamma} \rangle      \right).
 \label{uagam}
\end{split}
\end{equation}
The first term on the right-hand side is symmetric in $\beta\gamma$; the second and the third are both antisymmetric and
describe the response to a \emph{uniform magnetic field}, ${\bf B}$. 
In particular, the second contribution has only valence-band components and
is related to the Berry curvature; the third is a cross-gap (CG) contribution that obeys the following Sternheimer 
equation,
\begin{equation}
\begin{split}
 & \left(   \hat{H}_{\bf k} + a \hat{P}_{\bf k} - \epsilon_{n {\bf k}} \right) | u^{\rm CG}_{n {\bf k},\beta \gamma} \rangle = \\
 & \quad - \hat{Q}_{\bf k} \left( \left\{ \partial_\gamma \hat{H}_{\bf k} , \partial_\alpha \hat{P}_{\bf k} \right\}   
   - \left\{ \partial_\alpha \hat{H}_{\bf k} , \partial_\gamma \hat{P}_{\bf k} \right\} \right) | u^{(0)}_{n {\bf k}} \rangle.
\label{stern-B}
\end{split}
\end{equation}
This corresponds precisely to the linear response of the wavefunctions to a uniform ${\bf B}$-field 
as derived in Ref.~\onlinecite{Essin}.

\subsection{Electric field}

The standard treatment of the electric-field perturbation 
is based on the long-wavelength limit of a scalar potential
perturbation.~\cite{Gonze}
Such an approach, which we shall discuss in Sec.~\ref{sec:scalar}, 
is appealing for its simplicity; however, when pushed to higher 
orders in ${\bf q}$, it has the disadvantage of limiting the
scopes of the theory to the longitudinal components of many 
dispersion-related tensors.
(The transverse components of the flexoelectric tensor, for example, require 
a \emph{current-density} response theory,~\cite{Cyrus} while the 
scalar potential is only sensitive to the \emph{charge-density} response.)

Instead, here we shall work in an electromagnetic gauge where the
scalar potential vanishes, and the electric field is provided by
a vector potential that is slowly varying over time,
\begin{equation}
\EE = -\partial_t {\bf A}.
\end{equation}
To achieve this goal, we need to establish a time-dependent framework, 
where the external perturbation (in this case, the vector potential discussed
in the previous subsection) is applied dynamically.

The adequate formalism to attack this problem is provided by first-order adiabatic perturbation 
theory, which relates the adiabatic wavefunctions, $| \delta n \rangle$, to the static response functions
$|\partial_\lambda n \rangle$  via a Sternheimer equation,
\begin{equation}
(\hat{H} + a \hat{P} - \epsilon_n) | \delta n \rangle
 = i |\partial_{\lambda} n \rangle.
 \label{stern_adiab}
\end{equation}
Here $|\partial_\lambda n \rangle$ and $| \delta n \rangle$ describe the first-order response to $\lambda$
and $\dot{\lambda}$, respectively.
In the context of the electric field response, this translates into
\begin{equation}
\begin{split}
 \left( \hat H^{(0)}_{\bf k+q} + a \hat{P}_{\bf k+q} - \epsilon_{m {\bf k}} \right) & | u^{\Ea}_{m {\bf k,q}} \rangle = \\
  -i | u^{A_\alpha}_{m {\bf k,q}} \rangle - &\hat{Q}_{\bf k+q} V^{\Ea}_{\bf q} | u^{(0)}_{n {\bf k}} \rangle,
  \label{stern_e}
 \end{split}
\end{equation} 
where we have incorporated charge self-consistency via 
the usual SCF potential contribution, $V^{\Ea}_{\bf q}$.
Remarkably, the ${\bf A}$-field response functions play now the role of
\emph{external perturbation} in the context of the ${\bf E}$-field response,
\begin{equation}
\hat{Q}_{\bf k+q} \hat{H}^{\Ea}_{\bf k, q} | u^{(0)}_{m {\bf k}} \rangle \rightarrow | i u^{A_\alpha}_{m {\bf k,q}} \rangle.
\end{equation}
This allows us to write the mixed derivative with respect to an electric field and
a second perturbation $\lambda$ as the following stationary functional of 
$| u^{\Ea}_{m {\bf k,q}} \rangle$ and $| u^{\lambda}_{m {\bf k,q}} \rangle$,
\begin{eqnarray}
E^{\Ea^* \lambda}_{\bf q}  &=& s \int_{\rm BZ} [d^3 k] \, \sum_m  E^{\Ea^* \lambda}_{m\bf k,q} \nonumber \\
         && + \frac{1}{2} \int \int K_{\bf q}({\bf r,r'}) 
           n_{\bf q}^{\Ea *}({\bf r}) n_{\bf q}^{\lambda}({\bf r}')  d^3 r d^3 r'
\label{mixede}
\end{eqnarray}
(we have neglected the nonvariational term, as it is generally
absent in the case of the $\EE$-field response), where 
\begin{eqnarray}
E^{\Ea^* \lambda}_{m\bf k,q} &=& \langle u^{\Ea}_{m {\bf k,q}} | 
        \left( \hat H^{(0)}_{\bf k+q} + a \hat{P}_{\bf k+q} - \epsilon_{m {\bf k}} \right) | u^{\lambda}_{m {\bf k,q}} \rangle  \nonumber \\
    && + \langle u^{\Ea}_{m {\bf k,q}} | \hat{Q}_{\bf k+q} H^{\lambda }_{\bf k, q}
     | u^{(0)}_{m {\bf k}} \rangle  
      +  \langle   i u^{A_\alpha}_{m {\bf k,q}} 
      | u^{\lambda}_{m {\bf k,q}} \rangle. \nn
\label{mixede_mk}
\end{eqnarray}

Note that Eq.~(\ref{mixede}) is \emph{not} a variational function of
the ${\bf A}$-field response wavefunctions, $| i u^{A_\alpha}_{m {\bf k,q}} \rangle$.
Consequently, when calculating ${\bf q}$-derivatives of $E^{\Ea^* \lambda}_{\bf q}$, one needs 
to explicitly derive the functions $| i u^{A_\alpha}_{m {\bf k,q}} \rangle$ as one would do
for a ``standard'' external potential operator (e.g., corresponding to a phonon or a ``metric''~\cite{metric}
perturbation, as in the flexoelectric case of Sec.~\ref{sec:flexo}).
Note also that at ${\bf q}=0$ the above formulas trivially reduce to the 
standard treatment of the uniform electric field perturbation,~\cite{Gonze/Lee} of which 
they constitute the desired generalization to arbitrary ${\bf q}$-vectors.

Before closing this subsection, it is interesting to verify where the 
variational formulas derived here stand compared to the existing treatment 
of the polarization response~\cite{Cyrus} via Eq.~(\ref{adiab}).
By imposing the stationary condition Eq.~(\ref{stern_e}) to Eq.~(\ref{mixede}),
we obtain the following \emph{nonstationary} formula for the polarization response,
\begin{equation}
\frac{d P_\alpha^{\bf q}}{d \lambda} = -\frac{2}{\Omega} E^{\Ea^* \lambda}_{\bf q} = -\frac{2s}{\Omega} \int_{\rm BZ} [d^3 k] \, 
  \sum_m \langle   i u^{A_\alpha}_{m {\bf k,q}}   | u^{\lambda}_{m {\bf k,q}} \rangle.
  \label{dpdlam}
\end{equation}
It is not difficult to show that Eq.~(\ref{dpdlam}) exactly matches Eq.~(\ref{adiab}).
One just needs to recall the relationship between the current-density operator and
the vector potential perturbation, Eq.~(\ref{jalp}), and the sum-over-states expression
of the adiabatic wavefunctions [a consequence of Eq.~(\ref{stern_adiab})],
\begin{equation}
| \delta u^\lambda_{m\bf k, q} \rangle = i \sum_{n \in {\rm unocc}} | u^{(0)}_{n\bf k+q} \rangle    
    \frac{    \langle  u^{(0)}_{n\bf k+q}     | u^\lambda_{m\bf k, q} \rangle } {  \epsilon_{n\bf k+q}  - \epsilon_{m\bf k}  }.
    \label{sos_adiab}
\end{equation}
To go from Eq.~(\ref{adiab}) to Eq.~(\ref{dpdlam}) it suffices then to incorporate
Eq.~(\ref{sos_adiab}) into Eq.~(\ref{adiab}), and subsequently move the
energy denominator and the factor of $i$ from the right ($\lambda$-response) to the 
left (${\bf A}$-response) matrix element.
This derivation shows that, apart from irrelevant differences in the notation,
Eq.~(\ref{adiab}) can be regarded as the nonstationary~\cite{Gonze/Lee} counterpart 
of the variational functional, Eq.~(\ref{mixede}).

\subsection{Scalar potential}

\label{sec:scalar}

A monochromatic scalar potential perturbation 
simply involves adding $\varphi e^{i {\bf q \cdot r}}$ to the
local electrostatic potential; thus, in the language of this work,
the external perturbation is the unity operator at any ${\bf q}$,
\begin{equation}
\hat{H}^{\varphi}_{\bf q} = \mathcal{Q} = -1,
\end{equation}
where $\mathcal{Q}$ is the electron charge.
The mixed derivative functional involving a scalar potential and
a second perturbation $\lambda$ then reads as (note, as in the electric
field case, the disappearance of the nonvariational term)
\begin{eqnarray}
E^{\varphi^* \lambda}_{\bf q}  &=& s \int_{\rm BZ} [d^3 k] \, \sum_m  E^{\varphi^* \lambda}_{m\bf k,q} \nonumber \\
         && + \frac{1}{2} \int \int K_{\bf q}({\bf r,r'}) 
           n_{\bf q}^{\varphi *}({\bf r}) n_{\bf q}^{\lambda }({\bf r}')  d^3 r d^3 r', 
\label{mixedv}
\end{eqnarray}
where 
\begin{eqnarray}
E^{\varphi^* \lambda}_{m\bf k,q} &=& \langle u^{\varphi}_{m {\bf k,q}} | 
        \left( \hat H^{(0)}_{\bf k+q} + a \hat{P}_{\bf k+q} - \epsilon_{m {\bf k}} \right) | u^{\lambda}_{m {\bf k,q}} \rangle  \nonumber \\
    && - \langle u^{(0)}_{m {\bf k}} | \hat{Q}_{\bf k+q} 
     | u^{\lambda}_{m {\bf k,q}}  \rangle  \nonumber \\
     && +  \langle u^{\varphi}_{m {\bf k,q}} | \hat{Q}_{\bf k+q}  \hat H^{\lambda }_{\bf k, q} 
      | u^{(0)}_{m {\bf k}} \rangle.
\label{mixedv_mk}
\end{eqnarray}  
Differentiation with respect to $| u^{\lambda}_{m {\bf k,q}} \rangle$ yields the Sternheimer equation
for the first-order wavefunctions,
\begin{equation}
\begin{split}
 \left( \hat H^{(0)}_{\bf k+q} + a \hat{P}_{\bf k+q} - \epsilon_{m {\bf k}} \right) & | u^{\varphi}_{m {\bf k,q}} \rangle = \\
  -\hat{Q}_{\bf k+q} & \left( -1 + \hat{V}_{\bf q}^\varphi  \right) | u^{(0)}_{m {\bf k}} \rangle  ,
\end{split}
\label{sternphi}
\end{equation} 
where $\hat{V}_{\bf q}^\varphi$ is, as usual, the SCF 
contribution to the perturbation.

As we mentioned earlier, the scalar potential response vanishes at ${\bf q}=0$, and
any mixed derivative involving $\varphi$ identically vanishes at ${\bf q}=0$ as well. 
(Note that the perturbation doesn't vanish at ${\bf q}=0$, as it is a constant 
equal to $-1$ at any value of the wavevector.)
At first order in ${\bf q}$ one
recovers the standard treatment of the uniform electric field,~\cite{Gonze}
with the following relationship between the corresponding first-order wave functions,
\begin{equation}
|u^{\E_\delta}_{m {\bf k}} \rangle = |i  u^{\varphi}_{m {\bf k},\delta} \rangle.
\label{evsphi}
\end{equation}
Then, by combining Eq.~(\ref{evsphi}) with our higher-order formula, Eq.~(\ref{e2gamdel}), one 
can obtain useful information about the dispersion of the charge-density response of the system to an arbitrary perturbation.

To see the relationship between the scalar potential and the first-order charge density, 
one can insert Eq.~(\ref{sternphi}) into Eq.~(\ref{mixedv}) to obtain
a nonstationary expression for the mixed derivative,
\begin{equation}
E^{\varphi^* \lambda}_{\bf q}  = -s \int_{\rm BZ} [d^3 k] \, \sum_m \langle u^{(0)}_{m {\bf k}}| \hat{Q}_{\bf k+q} 
     |  u^{\lambda}_{m {\bf k,q}} \rangle,
\end{equation}     
which provides a direct link to the electronic contribution to the charge-density response,
\begin{equation}
\bar{\rho}_{\rm el,\bf q}^\lambda = - \frac{1}{\Omega} \int_\Omega d^3 r \, n^\lambda_{\bf q} ({\bf r}) = \frac{2}{\Omega} E^{\varphi^* \lambda}_{\bf q}.
\end{equation}
Note that $\bar{\rho}_{\rm el,\bf q}^\lambda$, the cell-averaged electronic charge density induced by the perturbation $\lambda$, 
differs from the cell average of $n^\lambda_{\bf q} ({\bf r})$ by a minus sign, which stems from the negative electron charge.

\subsection{Relationship to the continuity equation}

The fact that $E^{\varphi^* \lambda}_{\bf q}$ and $-E^{\Ea^* \lambda}_{\bf q}$ correspond,
respectively (modulo a factor of $2/\Omega$), to the charge-density and polarization response to 
the perturbation $\lambda$ implies that they must satisfy the continuity equation, 
$\bm{\nabla} \cdot {\bf P} = -\rho$.
In reciprocal space, this means that the following must be true,
\begin{equation}
i \sum_\alpha q_\alpha E^{\Ea^* \lambda}_{\bf q} = E^{\varphi^* \lambda}_{\bf q}.
\label{finiteq}
\end{equation}
The correctness of this result can be, of course, verified at the level of 
the finite-${\bf q}$ functionals, respectively Eq.~(\ref{mixede}) and Eq.~(\ref{mixedv}).
In the context of the present work, however, it is perhaps more insightful to 
verify Eq.~(\ref{finiteq}) in the long-wave limit, and use it as a ``sanity check'' 
of the formalism.
At the lowest orders in ${\bf q}$, Eq.~(\ref{finiteq}) leads to the following relationships,
\begin{eqnarray}
-E^{\Ea^* \lambda} &=& i E^{\varphi^* \lambda}_\alpha, \label{e0_v1} \\
i E^{\Ea^* \lambda}_\beta + i E^{\Eb^* \lambda}_\alpha &=& -E^{\varphi^* \lambda}_{\alpha \beta}.
\label{e1_v2}
\end{eqnarray}
(We have chosen the prefactors in such a way that all quantities are real numbers, and
that they match the sign conventions of Ref.~\onlinecite{artlin}.)
Eq.~(\ref{e0_v1}) is trivial to verify by using Eq.~(\ref{mixede}) and the Hellmann-Feynman theorem
applied to the first ${\bf q}$-gradient of Eq.~(\ref{mixedv}).
Eq.~(\ref{e1_v2}) can be checked by applying the higher-order formula Eq.~(\ref{e2gamdel}) to Eq.~(\ref{mixedv}), 
and by using the relationship existing between the scalar potential and the electric 
field response functions, Eq.~(\ref{evsphi}).
In the special case of $\lambda$ being an atomic displacement,
one can recognize the relationships between the multipolar expansion of the charge-density
and polarization response as established in Ref.~\onlinecite{artlin}.

\section{Dynamical quadrupole tensor}

\label{sec:quad}

\subsection{Theory}

Following the notation of Ref.~\onlinecite{artlin}, 
we can define the cell-integrated charge response to a
monochromatic atomic displacement as
\begin{equation}
Q^{\bf q}_{\kappa \beta} = \Omega \bar{\rho}_{\rm \bf q}^{\tau_{\kappa \beta}} = -i q_\beta Z_\kappa + 
         2   E^{\varphi^* \tau_{\kappa \beta} }_{\bf q},
\end{equation}
where $Z_\kappa$ is the pseudopotential charge, and $2E^{\varphi^* \tau_{\kappa \beta} }_{\bf q}$ is the mixed 
derivative of the energy with respect to a scalar potential [see Sec.~\ref{sec:scalar}, Eq.~(\ref{mixedv}) in 
particular] and an atomic displacement pattern of the type
\begin{equation}
{\bf R}_{l\kappa} = {\bf R}_{l\kappa}^0 + \bm{\tau}_{\kappa} e^{i {\bf q\cdot } {\bf R}_{l\kappa}^0}.
\label{phonon}
\end{equation}
($l$ and $\kappa$ are cell and sublattice indices, respectively; ${\bf R}_{l\kappa}^0$ indicates the unperturbed
atomic position; note that this perturbation differs from the standard implementations of DFPT~\cite{Baroni-01,Gonze} by
a phase factor, see Appendix~\ref{qderHatdis} for details.)

In the long-wave limit, $Q^{\bf q}_{\kappa \beta}$ can be written as a multipole expansion of the
charge density induced by an atomic displacement,
\begin{equation}
Q^{\bf q}_{\kappa \beta} = -i q_\gamma Q^{(1,\gamma)}_{\kappa \beta} - \frac{q_\gamma q_\delta}{2} Q^{(2,\gamma \delta)}_{\kappa \beta} + \cdots,
\end{equation}
where the dots stand for higher-order terms that we will not discuss in this work.
The first-order term corresponds to the Born effective charge tensor,
\begin{equation}
Z^*_{\kappa,\beta \gamma} = Q^{(1,\gamma)}_{\kappa \beta} = \delta_{\beta \gamma} Z_\kappa + \Delta Z_{\kappa,\beta \gamma},
\end{equation}
where the electronic contribution reads as 
\begin{eqnarray}
\Delta Z_{\kappa,\beta \gamma} = &=& 
   2 i s \int_{\rm BZ} [d^3 k] \, \sum_m \langle u^{(0)}_{m {\bf k}}  | \partial_\gamma \hat{P}_{\bf k} 
     |u^{\tau_{\kappa \beta}}_{m {\bf k}}  \rangle \nonumber \\
     &=& 2 i s \int_{\rm BZ} [d^3 k] \, \sum_m \langle \tilde{\partial}_\gamma u^{(0)}_{m {\bf k}}
     | u^{\tau_{\kappa \beta}}_{m {\bf k}}   \rangle \nonumber \\
   &=& -2 s \int_{\rm BZ} [d^3 k] \, \sum_m \langle  i \tilde{\partial}_\gamma u^{(0)}_{m {\bf k}} 
     | u^{\tau_{\kappa \beta}}_{m {\bf k}}  \rangle,
   \label{deltaz}  
\end{eqnarray}
thus recovering the already established result.~\cite{Baroni-01,Gonze/Lee} [We have applied the Hellmann Feynman theorem 
to the ${\bf q}$ derivative of the functional of Eq.~(\ref{mixedv}), combined with the fact
that the $\varphi$-response wavefunctions vanish at ${\bf q}=0$.]

The quadrupole tensor elements can be written as the second ${\bf q}$-gradients of $Q^{\bf q}_{\kappa \beta}$, 
\begin{equation}
Q_{\kappa\beta}^{(2,\gamma\delta)}= -2 E_{\gamma\delta}^{\varphi^* \tau_{\kappa\beta}}.
\label{eq_main_qdrpl}
\end{equation}
By using Eq.~(\ref{e1_v2}) we arrive at the following expression,
\begin{equation}
 E_{\gamma\delta}^{\varphi^* \tau_{\kappa\beta}}= -i E_{\gamma}^{\E_\delta^* \tau_{\kappa\beta}}
 -i E_{\delta}^{\E_\gamma^* \tau_{\kappa\beta}}.
 \label{symmquad}
\end{equation}
The first ${\bf q}$-gradient of the mixed response to an electric
field and to an atomic displacement can then be calculated by applying
Eq.~(\ref{e2gamma}) and Eq.~(\ref{e2gamma_mk}) to Eq.~(\ref{mixede}) and 
Eq.~(\ref{mixede_mk}), respectively,
\begin{equation}
\begin{split}
 & E_{\gamma}^{\E_\delta^* \tau_{\kappa\beta}} = s \int_{\rm BZ} [d^3k] \sum_{m} 
 E_{m{\bf k},\gamma}^{\E_\delta^* \tau_{\kappa\beta}}  + \\
 & \quad  \frac{1}{2} \int_{\Omega} \int K_{\gamma}({\bf r},{\bf r}') n^{\E_\delta} ({\bf r})
 n^{\tau_{\kappa\beta}}({\bf r}') d^3r d^3r',
 \end{split}
\end{equation}
\noindent with
\begin{equation}
\begin{split}
 & E_{m{\bf k},\gamma}^{\E_\delta^* \tau_{\kappa\beta}} = 
 \langle u_{m{\bf k}}^{\E_{\delta}} | \partial_{\gamma} \hat{H}^{(0)}_{{\bf k}} | u_{m{\bf k}}^{\tau_{\kappa\beta}} \rangle + \\
 & \quad \langle u_{m{\bf k}}^{\E_{\delta}}| \partial_{\gamma} \hat{Q}_{{\bf k}}  \hat{\mathcal{H}}_{{\bf k}}^{\tau_{\kappa\beta}}  | 
  u_{m{\bf k}}^{(0)} \rangle +
  \langle u_{m{\bf k}}^{(0)} |  V^{\E_{\delta}} \partial_{\gamma} \hat{Q}_{{\bf k}}| u_{m{\bf k}}^{\tau_{\kappa\beta}} \rangle +\\
 & \quad \langle  u_{m{\bf k}}^{\E_{\delta}}|  \hat{H}_{{\bf k},\gamma}^{\tau_{\kappa\beta}}  | u_{m{\bf k}}^{(0)} \rangle + 
 \langle i\, u_{m{\bf k},\gamma}^{A_{\delta}} |  u_{m{\bf k}}^{\tau_{\kappa\beta}} \rangle .
 \end{split}
 \label{qdrp_bks}
\end{equation}
Since the response functions need to be symmetrized according to Eq.~(\ref{symmquad}), we can
simplify the expression of $|u_{m{\bf k},\gamma}^{A_{\delta}} \rangle$, Eq.~(\ref{uagam}), and set 
\begin{equation}
\langle i\, u_{m{\bf k},\gamma}^{A_{\delta}} |  u_{m{\bf k}}^{\tau_{\kappa\beta}} \rangle \rightarrow 
  -\frac{i}{2}  \langle u_{m{\bf k}}^{(0)}| \partial_{\gamma\delta} \hat{P}_{{\bf k}} |  u_{m{\bf k}}^{\tau_{\kappa\beta}} \rangle,
  \label{symm_a2}
\end{equation}  
where $\partial_{\gamma\delta} \hat{P}_{{\bf k}}$ is the second ${\bf k}$-gradient of the valence-band projector
described in Appendix~\ref{app:a}.
[Other terms in Eq.~(\ref{uagam}) vanish as they are antisymmetric in the two indices.]
The explicit formula for the ${\bf q}$-derivative of the atomic displacement perturbation, 
$\hat{H}_{{\bf k},\gamma}^{\tau_{\kappa\beta}}$, is reported in Appendix ~\ref{qderHatdis}. 

Note that we could have obtained the exact same result by directly applying the higher-order formula 
Eq.~(\ref{e2gamdel}) to Eq.~(\ref{mixedv}), instead of using Eq.~(\ref{symmquad}).
The above procedure has the advantage that the intermediate quantity 
$E_{m{\bf k},\gamma}^{\E_\delta^* \tau_{\kappa\beta}}$ has also a well-defined physical meaning, 
as it relates to the ${\bf P}^{(1)}$-tensors discussed in Refs.~\onlinecite{artlin,chapter-15},
\begin{equation}
P^{(1,\gamma)}_{\delta,\kappa\beta} = \frac{2i}{\Omega} E_{\gamma}^{\E_\delta^* \tau_{\kappa\beta}}.
\end{equation}
These quantities can be interpreted as the first spatial moment of the polarization field induced by
an atomic displacement, and are required for the calculation of the so-called ``mixed'' contribution
to the flexoelectric tensor. 
Their in-depth discussion would bring us out of our main topic, and we defer it to a 
forthcoming publication.

\subsection{Computational parameters}
\label{sec:quad_np}

The computation of the quadrupole 
tensor has been implemented in the ABINIT~\cite{abinit} package as a \emph{postprocessing} of the DFPT response functions calculation. 
All the numerical results have been obtained employing Troullier-Martins norm conserving pseudopotentials and the Perdew-Wang~\cite{perdew/wang:1992} parametrization of the LDA. 
For our calculations on bulk Si, we use the calculated cell parameter of $a_0=$10.102 Bohr and two different crystal cells: (i) the
primitive 2-atoms cell, sampled with a Monkhorst-Pack (MP) mesh of $12 \times 12 \times 12$ ${\bf k}$-points, and (ii) a non-primitive 
6-atoms hexagonal cell with the translation vectors oriented along $[0 1 \bar{1}]$, $[1 0 \bar{1}]$ and $[111]$, sampled
with a $\Gamma$-centered $22 \times 22 \times 22$ ${\bf k}$-mesh.
We use a plane-wave cut-off of 20 Ha in both cell types.
We have also performed a convergence study by repeating our calculations at several different values 
of the cutoff and ${\bf k}$-point mesh resolution.
Regarding our calculations of ferroelectric PbTiO$_3$, we use a tetragonal 5-atoms unit cell,
with a plane-wave cut-off of 70 Ha and a $8\times 8 \times 8$ MP
mesh of ${\bf k}$-points.
We relax the unit cell until the forces are smaller than  $1.\times 10^{-6}$ Ha/bohr,
obtaining an aspect ratio of $c/a=1.046$ ($a=7.275$ bohr) and a spontaneous polarization 
$P_{\rm S}=0.78$ C/m$^2$.
These structural data are in excellent agreement with earlier calculations of the same system.~\cite{fixedd}

\subsection{Numerical results}

\begin{figure} 
\centering
\includegraphics[width=1.\columnwidth]{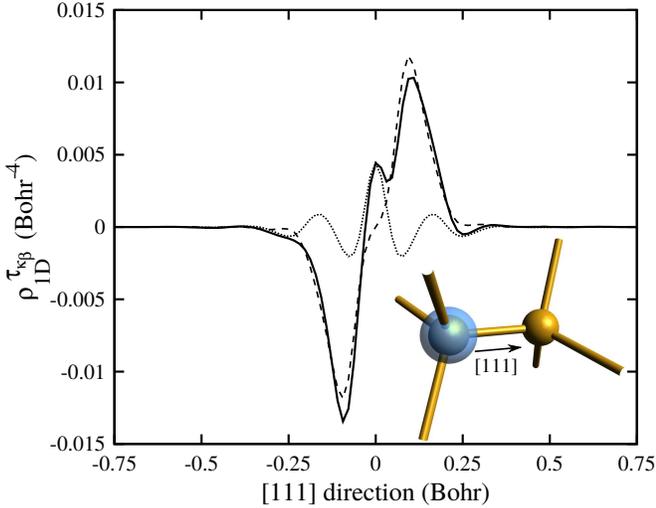}
\caption{Electron-density response to an atomic displacement along the [111] direction in bulk Si (solid line). Its symmetric (dotted line) and antisymmetric (dashed line) parts are also plotted. The first-order density has been averaged in-plane. The origin of the abscissas coincides with  
the position of the atomic sublattice highlighted in the inset.}
\label{fig_dens1}
\end{figure}

We first study bulk Si as a testcase. The quadrupolar tensor is defined
by a single material constant, $Q$, via the following expression,
\begin{equation}
Q_{\kappa\beta}^{(2,\gamma\delta)}= (-1)^{\kappa+1}\,Q\,|\varepsilon_{\beta\gamma\delta}|,
\end{equation}
$\varepsilon_{\beta\gamma\delta}$ is the Levi-Civita tensor. 
In order to benchmark the formalism, we first perform a calculation of $Q$ via
an independent real-space method, which does not rely on Eq.~(\ref{eq_main_qdrpl}).
To this end, we calculate the charge-density response to an atomic displacement 
along $[111]$, by using a Brillouin-zone unfolding procedure~\cite{sge} applied to the 6-atoms hexagonal cell.
In particular, we consider a stripe of 22 
equidistant {\bf q}-points ({\bf q}=0 included), spanning the entire Brillouin zone 
along the crystallographic $[111]$ direction, and calculate the first-order
densities associated with a phonon perturbation at each ${\bf q}$. (In practice, the number of independent ${\bf q}$-points reduces to 12 due to time-reversal symmetry.)
After unfolding, we readily obtain the induced charge density that corresponds 
to a displacement of an isolated atom.
(In practice, this approach corresponds to studying the displacement of
a plane of atoms in a supercell where the hexagonal unit is repeated 22 times along [111].)
 We report the plane averages of the first-order density in Fig.~\ref{fig_dens1}, where 
 we also show its decomposition into the antisymmetric and symmetric contributions. 
 A fast decay of the induced density is clearly observed, which allows us to 
 calculate the desired real-space moments with high numerical accuracy.
 The dipole moment correctly reproduces the pseudopotential charge, as expected.
The second real-space moment of the first-order charge is then related to 
$Q$ via
\begin{equation}
 Q^{\rm rs} = \frac{\sqrt{3}}{2} \epsilon_\infty \bar{Q}_{[111]}^{(2)},
 \label{qdrp_cdr}
 \end{equation}
 where the superscript rs stands for ``real space'', and $\epsilon_\infty$
 is the calculated electronic dielectric constant.

 Our calculated values are $\bar{Q}_{[111]}^{(2)}=1.178$ and $\epsilon_\infty=13.103$, which via Eq.~(\ref{qdrp_cdr}) yield a value of $Q^{\rm{rs}}=13.367$ 
 $\rm{e \cdot Bohr}$.
 With an equivalent choice of the computational parameters, by using
 our new method, Eq.~(\ref{eq_main_qdrpl}) and the primitive 2-atom cell 
 we obtain $Q=13.368$ $\rm{e \cdot Bohr}$. 
 The matching between the two approaches is essentially perfect, which
 demonstrates the soundness of our implementation.
 
\begin{figure*} 
\centering
\includegraphics[width=1.\textwidth]{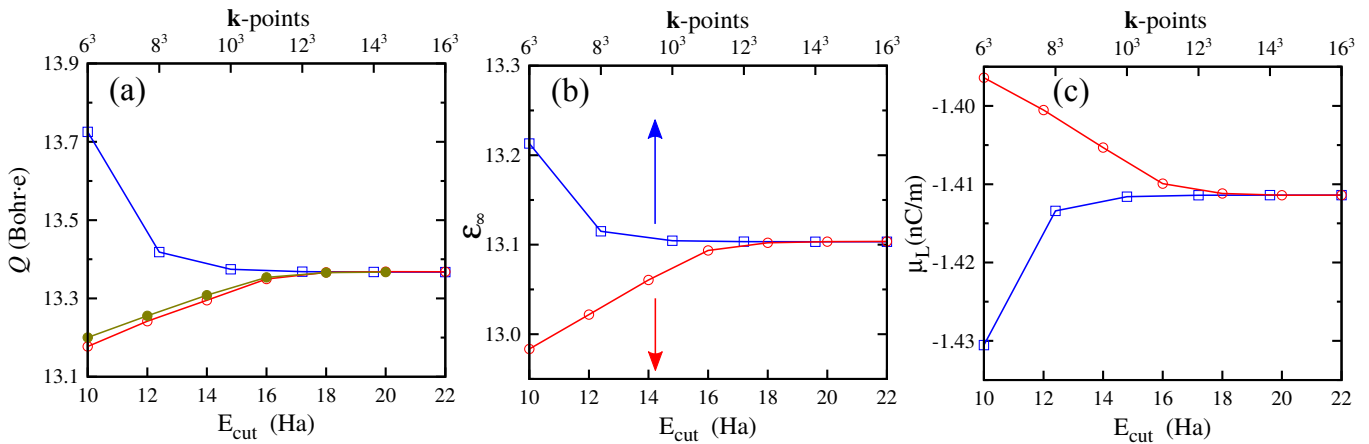}
\caption{Convergence of selected linear-response quantities with respect to the plane-wave cutoff and the density of the ${\bf k}$-point mesh.
Calculated dynamic quadrupole moment (a),  dielectric constant (b) and longitudinal flexoelectric coefficient (c) are shown. Empty blue squares (lines are a guide to the eye) are obtained by varying the plane-waves energy cutoff while keeping the $12\times 12\times 12$ \textbf{k}-points grid fixed. Red empty circles correspond to varying the \textbf{k}-points mesh resolution with a fixed energy cut-off of 20 Ha. Panel (a) includes the energy cut-off dependence of the quadrupole constant as calculated from the second moment of the induced charge density response (green circles).}
\label{fig_convergence}
\end{figure*}

 To better illustrate the plane-wave and ${\bf k}$-point mesh requirements of our new method, we have also performed a convergence study, where we compare the behavior of the quadrupole 
 tensor components (alongside with the flexoelectric response, which we will comment on in Section~\ref{sec:flexo}) to that of a ``standard'' linear-response quantity, the electronic dielectric constant. 
 The numerical results are plotted in Fig.~\ref{fig_convergence}(a-b) as a function of the plane-waves energy cutoff and the 
 number of \textbf{k}-points employed to sample the Brillouin zone. 
  One can clearly appreciate from the figure that the quadrupoles (panel a) and the dielectric constant (panel b) 
 converge equally fast with respect to both computational parameters.
 Moreover,  the agreement between $Q$ and $Q^{\rm rs}$ becomes better and better as the energy cutoff is increased.
 Both observations concur to put our new method, based on Eq.~(\ref{eq_main_qdrpl}), on very firm grounds. 
 Note that the calculation via Eq.~(\ref{eq_main_qdrpl}) is about an order of magnitude more efficient than the 
 alternative real-space method, as the latter requires calculating the phonon response at many ${\bf q}$-points, 
 while the former only requires $\Gamma$-point response functions as prerequisites.

\begin{table} 
\begin{center}
\begin{tabular}{c|rrrrr}\hline\hline
   & $\kappa=$Pb & $\kappa=$Ti & $\kappa=$O$_1$ & $\kappa=$O$_2$ & $\kappa=$O$_3$\\\hline
 $Q_{\kappa 3}^{(2,11)}$ & 2.264 & $-$3.545 & 2.884 & $-$4.186 & 0.406 \\
 $Q_{\kappa 3}^{(2,22)}$ & 2.264 & $-$3.545 & $-$4.186 & 2.884 & 0.406 \\
 $Q_{\kappa 1}^{(2,31)}$ & $-$0.062 & $-$3.799 & 3.123 & $-$1.115 & $-$1.784 \\
 $Q_{\kappa 2}^{(2,32)}$ & $-$0.062 & $-$3.799 & $-$1.115 & 3.123 & $-$1.784 \\
 $Q_{\kappa 3}^{(2,33)}$ & 1.240 & $-$0.195 & 2.027 & 2.027 & 6.653 \\\hline \hline 
\end{tabular}
\end{center}
\caption{ Quadrupole moments (in e$\cdot$Bohr) of PbTiO$_3$ calculated via Eq.~(\ref{eq_piezo}).
 Note that $Q_{\kappa \alpha}^{(2,\beta \gamma)} = Q_{\kappa \alpha}^{(2,\gamma\beta )}$.
 } 
\label{Tab_PTO_qdrp}
\end{table}

\begin{table}[b!]
\begin{center}
\begin{tabular}{c|ccc}  \hline\hline
   & $e_{113}=e_{223}$ & $e_{311}=e_{322}$ & $e_{333}$ \\\hline
Strain & 0.1547 & 0.3617 & $-$0.8345 \\
Quadrupoles & 0.1548 & 0.3614 & $-$0.8347 \\ \hline
Ref.~\onlinecite{Saghi-Szabo1998} & \emph{0.20} \hspace{6pt} & \emph{0.35} \hspace{6pt} & $-$\emph{0.88} \hspace{6pt}  \\\hline \hline 
\end{tabular}
\end{center}
\caption{Clamped-ion piezoelectric coefficients (in C/m$^2$) of PbTiO$_3$ calculated via two different methods. ``Strain'': Standard DFPT approach, relying on the strain~\cite{hamann-metric} 
and electric-field response.
``Quadrupoles'': From the quadrupole moments via Eq.~(\ref{eq_piezo}). 
Literature values from Ref.~\onlinecite{Saghi-Szabo1998} are shown in italics for comparison.}
\label{Tab_piezo}
\end{table}

Next, as a more ambitious test of our method,
we carry out a numerical verification
of Martin's formula~\cite{Martin} 
\begin{equation}
 e_{\alpha\beta\gamma}=-\frac{1}{2\Omega}\sum_{\kappa} 
  \left( Q_{\kappa\beta}^{(2,\alpha\gamma)} - Q_{\kappa\alpha}^{(2,\gamma\beta)} + Q_{\kappa\gamma}^{(2,\beta\alpha)} \right),
 \label{eq_piezo}
\end{equation}
relating the proper~\cite{vanderbilt:2000} 
clamped-ion piezoelectric tensor $e_{\alpha\beta\gamma}$ to the
sublattice sum of the dynamical quadrupoles. 
[Here, the first subscript ($\alpha$) of the piezoelectric tensor in the left-hand side 
indicates the polarization direction, whereas the other two indices ($\beta\gamma$) refer to
the strain tensor components.] 
In particular, we shall benchmark the value of $e_{\alpha\beta\gamma}$ 
computed from Eq.~(\ref{eq_piezo}) via the quadrupoles, 
against its value obtained as the mixed derivative 
of the energy with respect to components of the strain and the electric field. The latter is 
a standard DFPT quantity that we obtain by means of the metric tensor formulation 
by Hamann \emph{et al.}~\cite{hamann-metric} as implemented in the ABINIT package.   

We focus on a well-known piezoelectric system, the tetragonal phase of PbTiO$_3$. 
The quadrupole moments of each atom in the unit cell are shown in Tab.~\ref{Tab_PTO_qdrp}. 
As for the three independent PbTiO$_3$ piezoelectric tensor elements, they are reported
in Tab.~\ref{Tab_piezo}. The comparison between the coefficients from the two methods demonstrates an exceptionally good agreement, which improves up to the fifth decimal digit by increasing the density of ${\bf k}$-points to $14\times 14\times 14$. 
Our results also qualitatively agree with those of Ref.~\onlinecite{Saghi-Szabo1998}, 
wherein a different set of lattice parameters, pseudopotentials and exchange-correlation functionals were employed.

\section{Flexoelectricity}

\label{sec:flexo}

\subsection{Theory}

From the point of view of atomistic calculations, flexoelectricity can be
decomposed into three distinct contributions:~\cite{artlin} lattice-mediated, mixed and electronic.
In principle, all three can be written, by using the formalism developed in this work,
in terms of few basic ingredients. These are the mixed response to an electric
field, atomic displacement or metric-wave perturbation taken at first or second order in ${\bf q}$.
We defer the detailed implementation and test of the full flexoelectric tensor to a forthcoming
publication, and focus here on the purely electronic response only.

The electronic flexoelectric tensor can be written as the second derivative with respect to ${\bf q}$
of the current-density that is adiabatically induced by a ``clamped-ion'' acoustic phonon perturbation,~\cite{Cyrus} 
i.e. to a displacement pattern of the type
\begin{equation}
{\bf R}_{l\kappa} = {\bf R}_{l\kappa}^0 + {\bf u} e^{i {\bf q\cdot } {\bf R}_{l\kappa}^0}.
\label{acphon}
\end{equation}
Note the absence of the basis index on the perturbation parameter; this implies that
all atoms in the primitive cell should be displaced simultaneously with equal amplitude, ${\bf u}$.
Thus, a calculation of the flexoelectric tensor can be, in principle, carried out by
regarding Eq.~(\ref{acphon}) as the sublattice sum of Eq.~(\ref{phonon}), which leads to
the following practical scheme.
First, one writes the polarization response to the displacement of an \emph{individual} 
sublattice at finite ${\bf q}$; then, a second-order expansion in the wavevector 
${\bf q}$ is performed; finally, the clamped-ion flexoelectric tensor is written as a 
sublattice sum of the result~\cite{artlin}. This was indeed the strategy adopted in 
Ref.~\onlinecite{Cyrus}. 

In the context of this work, however, such an approach is 
impractical -- the phonon perturbation of Eq.~(\ref{phonon}) does not vanish
in the ${\bf q=0}$ limit. Therefore, Eq.~(\ref{e2gamdel}) cannot be 
directly applied to calculate expansion to second order in ${\bf q}$
of the corresponding polarization response.
To work around this obstacle, we shall follow Refs.~\onlinecite{metric,StengelPRB18} and
recast the acoustic phonon as a ``metric wave'' perturbation by operating
a coordinate transformation to the curvilinear co-moving frame.
We shall then write the polarization response to the acoustic phonon at finite 
${\bf q}$ as (following the notation of Ref.~\onlinecite{metric})
\begin{equation}
\overline{P}_{\alpha,\beta}^{'\bf q} = -\frac{2}{\Omega} E_{\bf q}^{\Ea^* (\beta)},
\label{dpdu}
\end{equation}
where $E_{\bf q}^{\Ea^* (\beta)}$ refers to the mixed derivative of Eq.~(\ref{mixede}) 
specialized to the case $\lambda=(\beta)$, and $(\beta)$ indicates a metric wave with the 
displacement field oriented along the Cartesian direction $\beta$. (The overline implies
cell averaging, and the prime indicates that we have implicitly 
discarded the magnetic-like contribution from rotation gradients, following 
the arguments of Refs.~\onlinecite{metric,StengelPRB18,Cyrus}.)

It is useful, at this stage, to recall~\cite{metric} two crucial properties of the metric wave: 
(i) both the perturbation and the response vanish at ${\bf q=0}$, 
\begin{equation}
\hat{H}^{(\beta)}_{\bf k, q=0} = 0, \qquad |u^{(\beta)}_{m\bf k,q=0}\rangle = 0;
\label{q0met}
\end{equation}
(ii) at first-order in ${\bf q}$, the metric wave reduces to the uniform 
strain perturbation, $\eta_{\beta \gamma}$, by Hamann \emph{et al.}~\cite{hamann-metric},
\begin{equation}
 \hat{H}_{{\bf k},\delta}^{(\beta)}= i \hat{H}_{{\bf k}}^{\eta_{\beta\delta}}, \qquad |u_{m{\bf k},\delta}^{(\beta)}\rangle= i |u_{m{\bf k}}^{\eta_{\beta\delta}}\rangle.
 \label{eq_metstr}
\end{equation}
As we shall see shortly, properties (i) and (ii) will allow us to write down a 
closed expression for the clamped-ion flexoelectric tensor by using the second-order 
formula, Eq.~(\ref{e2gamdel}).
Before doing that, it is useful to perform a consistency check of Eq.~(\ref{dpdu})
by showing that it correctly recovers the piezoelectric tensor at first order in ${\bf q}$
The clamped-ion piezoelectric tensor can be defined as
\begin{equation}
e_{\alpha \beta \gamma} = -i \frac{d}{d q_\gamma} \left(  \frac{d P_\alpha^{\bf q}}{d u_\beta}  \right) \Big|_{\bf q=0} = i \frac{2}{\Omega} E_{\gamma}^{\Ea^* (\beta)}.
\end{equation}
By applying Hellmann-Feynman theorem to Eq.~(\ref{mixede}) and by using Eq.~(\ref{q0met})
and Eq.~(\ref{eq_metstr}) we readily obtain
\begin{eqnarray}
e_{\alpha \beta \gamma} &=& i \frac{2s}{\Omega}  \int_{\rm BZ} [d^3 k] \, \sum_m \langle u^{\Ea}_{m {\bf k}} |  \hat{H}^{(\beta) }_{\bf k,\gamma} | u^{(0)}_{m {\bf k}} \rangle \nn
                        &=& - \frac{2s}{\Omega}  \int_{\rm BZ} [d^3 k] \, \sum_m \langle u^{\Ea}_{m {\bf k}} | \hat{H}_{{\bf k}}^{\eta_{\beta \gamma}} | u^{(0)}_{m {\bf k}} \rangle,
\end{eqnarray}
which matches the established result.~\cite{Gonze/Lee,hamann-metric}

The type-I clamped-ion flexoelectric tensor can now be written in terms of the following formula,
\begin{equation}
\mu^{\rm I}_{\alpha\beta,\gamma\delta} 
   =  \frac{1}{\Omega} E^{\E_\alpha^* (\beta)}_{\gamma\delta},
\end{equation}
where the mixed derivative is, as above, taken with respect to an electric field and the metric-wave perturbation.
By taking again into account the relationships existing between the metric $(\beta)$ and the strain perturbations $\eta_{\beta\delta}$,
the formulas for the second gradient , Eq.~(\ref{e2gamdel}) and Eq.~(\ref{e2gamdel_mk}), are as follow,

\begin{equation}
\begin{split}
  & \widetilde{E}^{\E^*_{\alpha} \, (\beta)}_{\gamma\delta} = s \int_{\rm BZ} [d^3k] \sum_{m}
  \widetilde{E}^{\E^*_{\alpha} \, (\beta)}_{m{\bf k},\gamma\delta} \\
  & \quad + \underbrace{
    \frac{i}{2} \int_{\Omega} \int K_{\gamma}({\bf r},{\bf r}') n^{\E_{\alpha}} ({\bf r})
 n^{\eta_{\beta\delta}}({\bf r}') d^3r d^3r' 
   }_{T_{\rm elst}}, 
\end{split}
  \label{eq_flexo_tens}
\end{equation}

\begin{equation}
\begin{split}
 & \widetilde{E}^{\E^*_{\alpha} \, (\beta)}_{m{\bf k},\gamma\delta}=
   \underbrace{
   i\langle u_{m{\bf k}}^{\E_{\alpha}} | \partial_{\gamma} \hat{H}^{(0)}_{{\bf k}} | u_{m{\bf k}}^{\eta_{\beta\delta}} \rangle 
   }_{T_1} \\ 
 & \quad +  \underbrace{
   i\langle u_{m{\bf k}}^{\E_{\alpha}} | \partial_{\gamma} \hat{Q}_{{\bf k}} \hat{\mathcal{H}}_{{\bf k}}^{\eta_{\beta\delta}} | u_{m{\bf k}}^{(0)} \rangle 
     }_{T_2} +  
    \underbrace{ 
 i\langle u_{m{\bf k}}^{(0)} | \hat{V}^{\E_{\alpha}} \partial_{\gamma} \hat{Q}_{{\bf k}} | u_{m{\bf k}}^{\eta_{\beta\delta}} \rangle 
    }_{T_3} \\ 
 & \quad + \underbrace{
 \frac{1}{2} \langle u_{m{\bf k}}^{\E_{\alpha}} | \hat{H}_{{\bf k},\gamma\delta}^{(\beta)} |  u_{m{\bf k}}^{(0)} \rangle 
   }_{T_4} +  
  \underbrace{
 i \langle i\, u_{m{\bf k},\gamma}^{A_{\alpha}} |  u_{m{\bf k}}^{\eta_{\beta\delta}} \rangle
   }_{T_5}.
 \label{eq_flexo_bks}
\end{split}
\end{equation}
Here we have labeled, for later reference, the five different terms of the band- and ${\bf k}$-resolved 
contribution as $T_{1-5}$, and the term deriving from the self-consistent energy via the gradient of the 
Coulomb kernel as $T_{\rm elst}$.
Most of the symbols are self-explanatory, as we have already encountered them in the
formula for the quadrupolar response.
Similarly to the quadrupole case, we shall use Eq.~(\ref{symm_a2}) to simplify
$T_5$; this is justified here because all our tests are performed on cubic materials,
where $T_5$ must be symmetric with respect to $\alpha \gamma$.
The formulas for the ${\bf q}$-derivatives of metric perturbation, $\hat{H}_{{\bf k},\gamma\delta}^{(\beta)}$, are elaborated in Appendix ~\ref{qderHmet}.

In the following subsection, we shall present our numerical results in type-II form by using 
\begin{equation}
\mu^{\rm II}_{\alpha\gamma,\beta \delta} = \mu^{\rm I}_{\alpha\beta, \gamma\delta} + \mu^{\rm I}_{\alpha\delta,\beta \gamma} - \mu^{\rm I}_{\alpha\gamma,\delta \beta }.
\label{type2}
\end{equation}
In practice, the transformation Eq.~(\ref{type2}) needs to be performed
explicitly only on $T_4$, since all the other terms are most naturally 
written in type-II form. (The explicit formula is reported in Appendix~\ref{qderHmet}.)
As we shall be dealing with cubic crystals only, we shall adopt the short-hand notation
$\mu_{\rm L}=\mu^{\rm II}_{11,11}$,  $\mu_{\rm T}=\mu^{\rm II}_{11,22}$ and $\mu_{\rm S}=\mu^{\rm II}_{12,12}$
for the three independent components, respectively: longitudinal (L), transverse (T) and shear(S).
We shall drop the ``II'' superscript and assume that the flexoelectric tensor is in type-II form
henceforth. 

\subsection{Computational parameters}

The computation of the clamped-ion flexoelectric 
tensor has also been implemented in the ABINIT~\cite{abinit} package.
The numerical results have been obtained with the same type of pseudopotentials and XC functional as in the Section~\ref{sec:quad_np}.
For our calculations on noble gas atoms He, Ar and Kr, we use a large cell of $14\times 14 \times 14$ a.u., with a plane-wave cut-off of 90 Ha and a $2 \times 2 \times 2$ ($4 \times 4 \times 4$) mesh of ${\bf k}$-points to sample the Brillouin zone of He and Ar (of Kr). 
For our calculations on SrTiO$_3$, we used a cubic 5-atoms unit cell with an optimized cell parameter of $a_0$=7.267 Bohr, with a plane-wave cut-off of 70 Ha and a $8 \times 8 \times 8$ mesh of  ${\bf k}$-points.
Regarding Si, we use the 2-atom primitive cell with the same computational parameters as described in the Section~\ref{sec:quad_np}. We have also performed a convergence study of the calculated Si flexoelectric tensor by varying the cut-off and ${\bf k}$-point mesh resolution.

\subsection{Numerical results}

\begin{table} 
\begin{center}
\begin{tabular}{c|rrr}\hline\hline
   & \multicolumn{1}{c}{$\mu_{\rm L}$} & \multicolumn{1}{c}{$\mu_{\rm T}$} &  \multicolumn{1}{c}{$\mu_{\rm S} \, (\times 10^{-4})$} \\\hline
He & $-$0.479 ($-$0.479$^a$) & $-$0.479 ($-$0.479$^a$) & $-$0.08 ($-$0.08$^a$) \\
Ar & $-$4.821 ($-$4.813$^a$) & $-$4.823 ($-$4.820$^a$) & $-$1 \hspace{7pt} ($-$10$^a$) \hspace{1pt} \\
Kr & $-$6.471 ($-$6.474$^a$) & $-$6.477 ($-$6.476$^a$) & $-$4 \hspace{7pt} ($-$20$^a$) \hspace{1pt} \\ \hline \hline 
\end{tabular}
\end{center}
\caption{Flexoelectric coefficients (in pC/m) of noble-gas atom systems. $^a$ Reference~\onlinecite{metric}. } 
\label{Tab_ng_flexo}
\end{table}

In order to test our method, we first study a simple cubic crystal lattice consisting of isolated 
noble-gas atoms, as already investigated in 
Refs.\onlinecite{artgr,chapter-15,StengelPRB18,Cyrus,metric}. This toy model presents the advantage 
that its flexoelectric coefficients can be determined analytically,
based on the macroscopic electric tensor and the second real-space moment of the unperturbed 
atomic charge. 
In particular, the three independent flexoelectric coefficients 
as calculated from a metric-wave perturbation, must fulfill the conditions~\cite{metric,StengelPRB18}  
\begin{equation}
\mu_{\rm L} = \mu_{\rm T}, \qquad \mu_{\rm S}=0.
\label{irc}
\end{equation}

\begin{table}[b]
\begin{center}
\begin{tabular}{c|rrrrrr}\hline\hline
          & \multicolumn{1}{c}{$T_{\rm elst}$} & \multicolumn{1}{c}{$T_1$} & \multicolumn{1}{c}{$T_2$} & \multicolumn{1}{c}{$T_3$} & \multicolumn{1}{c}{$T_4$} & \multicolumn{1}{c}{$T_5$} \\\hline
$\mu_{\rm L}$ & $-$6.472	& 3.512 &	$-$0.257	 & 1.909	& $-$9.367	& 5.854\\
$\mu_{\rm T}$ & $-$1.885	&$-$2.286 &	$-$0.257	 & 0.000	& $-$0.395	& 0.000 \\
$\mu_{\rm S}$ & $-$1.885	& 2.907 &	0.000	 & 0.954	& $-$4.903	& 2.926 \\\hline \hline 
\end{tabular}
\end{center}
\caption{Contribution to the Ar flexoelectric coefficients (in pC/m) from the different terms of Eqs.~(\ref{eq_flexo_tens}) and~(\ref{eq_flexo_bks}).} 
\label{Tab_Ar_contr}
\end{table} 

Tab.~\ref{Tab_ng_flexo} shows the flexoelectric coefficients calculated for He, Ar and Kr. 
It is clear from the reported data that the expected relationships, Eq.~(\ref{irc}), are satisfied to a high degree of accuracy, and our coefficients are in excellent agreement with those obtained in previous works.~\cite{metric} 
The largest deviation is shown by Kr: being it a larger atom, the
overlap between neighboring images is likely to be more pronounced than in the
other cases, which justifies the discrepancies we observe with
respect to the expectations of the isolated atom model.

At this stage, it is worth emphasizing that such test is by no means trivial; on the contrary, it represents a very stringent benchmark for our formalism.
To demonstrate this point, in Tab.~\ref{Tab_Ar_contr} we show a breakdown of  the three independent coefficients of the Ar-based crystal into the contributions of the individual terms appearing in Eqs.~(\ref{eq_flexo_tens}) and~(\ref{eq_flexo_bks}).
The data in the table show a much more complex behavior than the final results of Tab.~\ref{Tab_ng_flexo} would suggest. 
In particular, the conditions $\mu_{\rm L}=\mu_{\rm T}$ and $\mu_{\rm S}=0$ are not fulfilled by \emph{any} of the individual terms 
(an exception is $T_2$, but it only contributes a tiny fraction of the final value); 
instead, the cancellation of the shear component and the equality between the transverse and longitudinal 
ones both result from a subtle balance between all the terms.
Since each term involves a different combination of the input response functions and of the 
perturbations, such an accurate compensation clearly demonstrates the robustness of the numerical 
implementation.

\begin{table} 
\begin{center}
\begin{tabular}{c|rrr}\hline\hline
   & \multicolumn{1}{c}{$\mu_{\rm L}$} & \multicolumn{1}{c}{$\mu_{\rm T}$} &\multicolumn{1}{c}{ $\mu_{\rm S}$} \\\hline
Si (this work) & $-$1.4114 & $-$1.0491 & $-$0.1895  \\
Ref.\onlinecite{metric} & $-$1.4110 & $-$1.0493 & $-$0.1894 \\ \hline
SrTiO$_3$ (this work) & $-$0.8848 & $-$0.8262 & $-$0.0823 \\
Ref.\onlinecite{metric} & $-$0.8851 & $-$0.8260 & $-$0.0823\\
Ref.\onlinecite{artcalc} & $-$0.883  \hspace{1.5pt} & $-$0.825 \hspace{1.5pt} & $-$0.082 \hspace{1.5pt} \\ \hline \hline 
\end{tabular}
\end{center}
\caption{Flexoelectric coefficients (in nC/m) of Si and SrTiO$_3$ along with previous values found in the literature.}
\label{Tab_cub_flexo}
\end{table}

We have also calculated the electronic contribution to the flexoelectric tensor of two real materials, Si and SrTiO$_3$. 
The converged values of the flexoelectric coefficients are shown in Tab.~\ref{Tab_cub_flexo}, where we also compare them to the relevant literature data~\cite{metric,artcalc,Cyrus}. 
Again, the excellent agreement with the published values is clear. 
Nevertheless, we stress that our results are obtained with a small fraction of the computational effort that was formerly needed.  

As a final benchmark, we study the convergence of the flexoelectric coefficients of Si as a function of the ${\bf k}$-point mesh resolution and of the plane-wave cutoff. The results for $\mu_{\rm L}$ are shown in Fig.~\ref{fig_convergence}(c). (The convergence of the two other independent components 
is qualitatively similar to the longitudinal one.)
Analogously to the case of the quadrupoles, the flexoelectric coefficients converge at the same rate as the dielectric tensor. This means that all the spatial dispersion properties that we have calculated in this 
work require a computational effort that is comparable to the study of 
other standard linear-response quantities, such as the electronic dielectric tensor.

\section{Conclusions and outlook}

\label{sec:outlook}

We have established a general method to perform a systematic study of
spatial dispersion effects in the framework of density-functional perturbation theory.
As a practical demonstration, we have implemented the dynamical quadrupole tensor and the 
clamped-ion flexoelectric tensor in the ABINIT package, and performed extensive numerical 
tests.
This work opens a number of exciting avenues for future research, which we shall briefly sketch hereafter.

First, we expect that the knowledge of the dynamical quadrupoles will allow for an improved
description of the interatomic force constants, thereby enabling a more accurate computation of 
the phonon band structures. 
This might be important for certain material classes, such as piezoelectrics, where 
the treatment of long-range electrostatics is crucial for reproducing the correct
sound velocity.~\cite{Wu-05}
On a different note, our theory 
might also prove itself very helpful 
in establishing higher-order multipolar generalizations~\cite{Benalcazar17} of the Berry-phase 
theory of polarization~\cite{king-smith/vanderbilt:1993}.
Indeed, our expressions for the dynamical quadrupole and flexoelectric tensors
can be regarded as the linear variation of the ``bulk 
quadrupolization''~\cite{Wheeler-arxiv} with respect to a zone-center lattice distortion 
or uniform strain, respectively.
There are intriguing parallels to the theory of multipolar magnetic orders~\cite{Ederer-07,Spaldin-13} 
as well, which will certanly stimulate further studies.

%

Second, the treatment of flexoelectric effects beyond the clamped-ion level should be
relatively straightforward by following the same guidelines as we did here. Both
the ``mixed'' and ``lattice-mediated'' contributions involve first or second
derivatives of the polarization response to a phonon, or the force-constant matrix,
just like the electronic contribution. These additional pieces involve similar 
formulas, only with a slightly different combination of the basic response functions
(electric field, atomic displacement or uniform strain).
Thus, the calculation of the full flexoelectric tensor for an insulating crystal 
or nanostructure of arbitrary symmetry looks now well within reach. We expect that,
once implemented, it will involve a computational effort that is comparable to the 
calculation  of the piezoelectric tensor.

Third, the method can be easily adapted to compute other spatial dispersion effects,
for example the natural optical rotation tensor. (The latter can be written as the 
first gradient with respect to the wavevector of the dielectric tensor.) First-principles
calculations of natural gyrotropy are starting to appear;~\cite{tellurium-ivo} 
we expect that, by bringing it within the scopes of DFPT, the formalism presented here
will greatly simplify the calculation of this interesting quantity as well.
In the context of ferroic materials, we also expect our method to facilitate the development 
of first-principles based continuum models~\cite{rotopolar} and effective Hamiltonians~\cite{heff},
where gradient-mediated couplings often play an important role.

More generally, our work has revealed a profound connection between spatial
dispersion and orbital magnetism that, in our opinion, deserves further attention. 
Whenever the polarization response to a perturbation is needed at first order in 
the wavevector ${\bf q}$, one of the contributions necessarily involves the 
wave-function response to a gradient of ${\bf A}$, and hence to a uniform \emph{magnetic} 
field.
This can be tentatively interpreted as a ``gyrotropic'' contribution to the 
response, and is only present in certain crystal classes; we were unable to discuss 
it here because of space limitations, but we regard it as yet another interesting
topic for future studies.

\begin{acknowledgments}

 We acknowledge the support of Ministerio de Economia,
 Industria y Competitividad (MINECO-Spain) through
 Grants  No.  MAT2016-77100-C2-2-P  and  No.  SEV-2015-0496,
 and  of Generalitat de Catalunya (Grant No. 2017 SGR1506).
 This project has received funding from the European
 Research Council (ERC) under the European Union's
 Horizon 2020 research and innovation program (Grant
 Agreement No. 724529). Part of the calculations were performed at
 the Supercomputing Center of Galicia (CESGA).
\end{acknowledgments}

\appendix

\begin{widetext}

\section{\label{app:a} Response to the electromagnetic vector potential in the long-wavelength limit}

The functions $| u^{A_\alpha}_{m {\bf k,\gamma}} \rangle$ are the ${\bf q}$-derivatives 
of the functions $|  u^{A_\alpha}_{m {\bf k,q}} \rangle$, which are defined as the solutions of the 
Sternheimer equation, Eq.~(\ref{stern-A}).
In the following, we shall proceed to explicitly demonstrate Eq.~(\ref{uagam}), by 
performing the formal expansion of Eq.~(\ref{stern-A}) to first order in ${\bf q}$.

By deriving both sides of the Sternheimer equation with respect to $q_\gamma$, one obtains
\begin{eqnarray}
\left(   \hat{H}^{(0)}_{\bf k} + a \hat{P}_{\bf k} - \epsilon^{(0)}_{n {\bf k}} \right) 
  | u^{A_\alpha}_{n {\bf k}, \gamma} \rangle &=& 
  -\left( \partial_\gamma \hat{H}^{(0)}_{\bf k}  + 
 a  \partial_\gamma \hat{P}_{\bf k}  \right) 
 | u^{A_\alpha}_{n {\bf k}} \rangle 
  -  \partial_\gamma \hat{Q}_{\bf k} \, \partial_\alpha \hat{H}^{(0)}_{\bf k} \,
 | u^{(0)}_{n {\bf k}} \rangle 
 - \frac{1}{2} \hat{Q}_{\bf k} \, \partial^2_{\alpha \gamma} \hat{H}^{(0)}_{\bf k} \, | u^{(0)}_{n {\bf k}} \rangle.
\end{eqnarray}
(We shall drop the superscript ``(0)'' on the ground-state Hamiltonian
operator from now on, to simplify the notation.)
One can now use the zero-th order result, Eq.~(\ref{ua0}), to achieve the following expression
(I shall also use $\partial_\alpha \hat{Q}_{\bf k}=-\partial_\alpha \hat{P}_{\bf k}$
whenever appropriate),
\begin{eqnarray}
\left(   \hat{H}_{\bf k} + a \hat{P}_{\bf k} - \epsilon_{n {\bf k}} \right) 
  | u^{A_\alpha}_{n {\bf k},  \gamma} \rangle &=& 
  \Big( \partial_\gamma \hat{H}_{\bf k} \, \partial_\alpha \hat{Q}_{\bf k}  - 
 a  \partial_\gamma \hat{P}_{\bf k} \, \partial_\alpha \hat{P}_{\bf k} 
 -  \partial_\gamma \hat{Q}_{\bf k} \, \partial_\alpha \hat{H}_{\bf k} \,
 - \frac{1}{2} \hat{Q}_{\bf k} \, \partial_\alpha \partial_\gamma \hat{H}_{\bf k} \Big) | u^{(0)}_{n {\bf k}} \rangle.
\end{eqnarray}
It is convenient, at this point, to separately treat the contributions that are 
symmetric and antisymmetric under $\alpha \gamma$ exchange.

\subsection{Symmetric part}

We have
\begin{eqnarray}
 \left(   \hat{H}_{\bf k} + a \hat{P}_{\bf k} - \epsilon_{n {\bf k}} \right) 
  \left( | u^{A_\alpha }_{n {\bf k}, \gamma} \rangle + | u^{A_\gamma}_{n {\bf k},\alpha } \rangle \right) &=& 
  \Big( - a  \partial_\gamma \hat{P}_{\bf k} \, \partial_\alpha \hat{P}_{\bf k}  
       - a  \partial_\alpha \hat{P}_{\bf k} \, \partial_\gamma \hat{P}_{\bf k}  
 + \partial_\gamma \hat{H}_{\bf k} \, \partial_\alpha \hat{Q}_{\bf k}   
   + \partial_\alpha \hat{H}_{\bf k} \, \partial_\gamma \hat{Q}_{\bf k}  \nonumber \\
 && \quad -  \partial_\gamma \hat{Q}_{\bf k} \, \partial_\alpha \hat{H}_{\bf k}  
   -  \partial_\alpha \hat{Q}_{\bf k} \, \partial_\gamma \hat{H}_{\bf k} 
  - \hat{Q}_{\bf k} \, \partial_\alpha \partial_\gamma \hat{H}_{\bf k} \Big) | u^{(0)}_{n {\bf k}} \rangle.
\end{eqnarray}
The long parenthesis on the right-hand side contains a total of seven terms.
The third to the seventh can be rewritten more compactly by observing that
$\partial^2_{\alpha \gamma} \left[ \hat{Q}_{\bf k},  \hat{H}_{\bf k} \right] = 0$, 
leading to
\begin{eqnarray}
 \left(   \hat{H}_{\bf k} + a \hat{P}_{\bf k} - \epsilon_{n {\bf k}} \right) 
   \left( |u^{A_\alpha }_{n {\bf k}, \gamma} \rangle + | u^{A_\gamma}_{n {\bf k},\alpha } \rangle \right) = 
  \Big(- a  \partial_\gamma \hat{P}_{\bf k} \, \partial_\alpha \hat{P}_{\bf k}  
       - a  \partial_\alpha \hat{P}_{\bf k} \, \partial_\gamma \hat{P}_{\bf k}  
 - \left[ \hat{H}_{\bf k}, \partial^2_{\alpha \gamma} \hat{Q}_{\bf k} \right]  \Big) | u^{(0)}_{n {\bf k}} \rangle.
\end{eqnarray}
The solution is given by
\begin{equation}
| u^{A_\alpha }_{n {\bf k}, \gamma} \rangle + | u^{A_\gamma}_{n {\bf k},\alpha } \rangle = 
-\partial^2_{\alpha \gamma} \hat{Q}_{\bf k} | u^{(0)}_{n {\bf k}} \rangle = 
\partial^2_{\alpha \gamma} \hat{P}_{\bf k} | u^{(0)}_{n {\bf k}} \rangle  .
\label{symm}
\end{equation}
To justify the above derivation, observe that
\begin{displaymath}
\partial^2_{\alpha \gamma} (PP) = 
\partial_\alpha P \partial_\gamma P + \partial_\gamma P \partial_\alpha P + \partial^2_{\alpha \gamma} P P +
P  \partial^2_{\alpha \gamma} P.
\end{displaymath}
Using the idempotency of $P$, this immediately leads to
\begin{displaymath}
P \, \partial^2_{\alpha \gamma} P \, P = -P \left( \partial_\gamma P  \, \partial_\alpha P + 
                                                             \partial_\alpha P \, \partial_\gamma P \right) P.
\end{displaymath}
Note that the response is purely ``geometric'', i.e. only given in terms of the 
ground-state wavefunctions. In spite of that, the response contains both valence
and conduction-band components, as the operator $\partial^2_{\alpha \gamma} \hat{P}_{\bf k}$
generally has both inner and cross-gap matrix elements. 

\subsection{Antisymmetric part}

The antisymmetric part can be written as follows,
\begin{eqnarray}
 \left(   \hat{H}_{\bf k} + a \hat{P}_{\bf k} - \epsilon_{n {\bf k}} \right) 
  \left( | u^{A_\alpha }_{n {\bf k}, \gamma} \rangle - | u^{A_\gamma}_{n {\bf k},\alpha } \rangle \right) &=& 
  \Big(- a  \partial_\gamma \hat{P}_{\bf k} \, \partial_\alpha \hat{P}_{\bf k}  
       + a  \partial_\alpha \hat{P}_{\bf k} \, \partial_\gamma \hat{P}_{\bf k}   
  + \partial_\gamma \hat{H}_{\bf k} \, \partial_\alpha \hat{Q}_{\bf k}   
    \nonumber \\
 && \quad - \partial_\alpha \hat{H}_{\bf k} \, \partial_\gamma \hat{Q}_{\bf k}  -  \partial_\gamma \hat{Q}_{\bf k} \, \partial_\alpha \hat{H}_{\bf k}  
   +  \partial_\alpha \hat{Q}_{\bf k} \, \partial_\gamma \hat{H}_{\bf k} \Big) | u^{(0)}_{n {\bf k}} \rangle. 
\end{eqnarray}
%
This can be expressed more compactly by using (anti)commutators,
\begin{eqnarray}
\left(   \hat{H}_{\bf k} + a \hat{P}_{\bf k} - \epsilon_{n {\bf k}} \right) 
  \left( | u^{A_\alpha }_{n {\bf k}, \gamma} \rangle - | u^{A_\gamma}_{n {\bf k},\alpha } \rangle \right) &=& 
  \Big(- a [ \partial_\gamma \hat{P}_{\bf k} , \partial_\alpha \hat{P}_{\bf k} ] 
   - \left\{ \partial_\gamma \hat{H}_{\bf k} , \partial_\alpha \hat{P}_{\bf k} \right\}   
   + \left\{ \partial_\alpha \hat{H}_{\bf k} , \partial_\gamma \hat{P}_{\bf k} \right\}\Big) | u^{(0)}_{n {\bf k}} \rangle. 
\end{eqnarray}
To recast the above equation into a more transparent form, it is useful to work 
out the following expression,
\begin{eqnarray}
\left[H, \partial_\gamma P \partial_\alpha P - \partial_\alpha P \partial_\gamma P \right] &=& 
 \left[H,\partial_\gamma P \right] \partial_\alpha P - \left[H,\partial_\alpha P \right] \partial_\gamma P
-\partial_\gamma P \left[\partial_\alpha P,H\right] + \partial_\alpha P \left[\partial_\gamma P,H\right]\nonumber \\ &=& 
-\left[\partial_\gamma H, P \right] \partial_\alpha P + \left[\partial_\alpha H, P \right] \partial_\gamma P
+\partial_\gamma P \left[ P,\partial_\alpha H\right] - \partial_\alpha P \left[ P,\partial_\gamma H\right]\nonumber \\ &=& 
 P \partial_\gamma H \partial_\alpha P - P \partial_\alpha H \partial_\gamma P 
-P \partial_\gamma P \partial_\alpha H + P \partial_\alpha P \partial_\gamma H\nonumber \\ &=& 
 P \left\{  \partial_\gamma H, \partial_\alpha P \right\} - P \left\{  \partial_\alpha H, \partial_\gamma P \right\},
\end{eqnarray}
where the fact that the expression must be applied to a valence ket has been used to go from the third to the
fourth line.
The Sternheimer equation can then be rewritten as
\begin{eqnarray}
 \left(   \hat{H}_{\bf k} + a \hat{P}_{\bf k} - \epsilon_{n {\bf k}} \right) 
   \left( | u^{A_\alpha }_{n {\bf k}, \gamma} \rangle - | u^{A_\gamma}_{n {\bf k},\alpha } \rangle \right) &=& 
 -\left( \hat{H}_{\bf k} + a \hat{P}_{\bf k} - \epsilon_{n {\bf k}} \right) 
  \left[ \partial_\gamma \hat{P}_{\bf k} , \partial_\alpha \hat{P}_{\bf k} \right] | u^{(0)}_{n {\bf k}} \rangle  \nonumber\\  
&&  
- \hat{Q}_{\bf k} \left( \left\{ \partial_\gamma \hat{H}_{\bf k} , \partial_\alpha \hat{P}_{\bf k} \right\}   
   + \left\{ \partial_\alpha \hat{H}_{\bf k} , \partial_\gamma \hat{P}_{\bf k} \right\} \right) | u^{(0)}_{n {\bf k}} \rangle. \nonumber
\end{eqnarray}
and, finally,
\begin{eqnarray}
&&\left(   \hat{H}_{\bf k} + a \hat{P}_{\bf k} - \epsilon_{n {\bf k}} \right)  | u^{\rm CG}_{n {\bf k}, \gamma} \rangle = 
 - \hat{Q}_{\bf k} \left( \left\{ \partial_\gamma \hat{H}_{\bf k} , \partial_\alpha \hat{P}_{\bf k} \right\}   
   - \left\{ \partial_\alpha \hat{H}_{\bf k} , \partial_\gamma \hat{P}_{\bf k} \right\} \right) | u^{(0)}_{n {\bf k}} \rangle,
\end{eqnarray}
where we have defined 
\begin{equation}
| u^{\rm CG}_{n {\bf k}, \gamma} \rangle = | u^{A_\alpha }_{n {\bf k}, \gamma} \rangle - | u^{A_\gamma}_{n {\bf k},\alpha } \rangle + 
 \left[ \partial_\gamma \hat{P}_{\bf k} , \partial_\alpha \hat{P}_{\bf k} \right] | u^{(0)}_{n {\bf k}} \rangle.
 \label{antisymm}
\end{equation}
By combining Eq.~(\ref{symm}) and Eq.~(\ref{antisymm}) we recover Eq.~(\ref{uagam}).


\section{${\bf q}$-derivatives of first order Hamiltonians}

\subsection{\label{qderHatdis}First ${\bf q}$-derivative of atomic displacement Hamiltonian}

The first-order Hamiltonian with respect to an atomic displacement consists of a local potential plus a nonlocal separable contribution,   

\begin{eqnarray}
V_{\bf q}^{{\rm loc},\tau_{\kappa \beta} }({\bf G}) &=& -i (G_\beta + q_\beta) \frac{1}{\Omega}  
e^{-i{\bf G} \cdot \bm{\tau}_\kappa} \, v^{\rm loc}_\kappa ({\bf G + q}), \label{vloc} \\
V^{{\rm sep},\tau_{\kappa \beta}}_{\bf k,q}({\bf G},{\bf G}') &=& -i(G_\beta + q_\beta - G'_\beta) 
\frac{1}{\Omega} \sum_{\mu} e^{-i({\bf G-G'})\cdot \bm{\tau}_\kappa} \,
e_{\mu \kappa} \zeta_{\mu \kappa}({\bf k +q + G}) \zeta^*_{\mu \kappa}({\bf k + G}').
\label{vsep}
\end{eqnarray}

\noindent Note that the above formulas slightly differ from the standard implementations of DFPT.~\cite{Gonze} 
This is because in formulating them we have used a different convention~\cite{artlin} for the sublattice-dependent phase factors.
Such difference is rooted in our assumption of a displacement pattern of the type
 $e^{i\bf{q}\bf{R}_{l\kappa}}$, instead of the typical one ($e^{i\bf{q}\bf{R}_l}$). 
 This leads to a much simpler and physically transparent treatment of the long-wave expansion. At ${\bf q}$=0, the present theory 
 reduces to the standard treatment.

By differentiating the above formulas, we arrive at the following expressions for the first ${\bf q}$-gradients (at ${\bf q}$=0) 
of the perturbation, which are necessary for the dynamical quadrupoles calculation,

\begin{eqnarray}
   V_{\gamma}^{{\rm loc},\tau_{\kappa\beta}}({\bf G}) &=&
 -i\frac{1}{\Omega} e^{-i {\bf G} \boldsymbol{\tau}_{\kappa}} 
 \Big( \delta_{\beta\gamma}  \, v_{\kappa}^{\rm loc}(G) + 
  \frac{G_{\beta}G_{\gamma}}{G} \, v_{\kappa}^{\rm loc}(G)' \Big), \\ \nonumber
   V_{{\bf k},\gamma}^{{\rm sep},\tau_{\kappa\beta}}({\bf G},{\bf G}')&=& 
  -i \frac{1}{\Omega} 
 \sum_{\mu} e^{-i ({\bf G}-{\bf G}') \boldsymbol{\tau}_{\kappa}} e_{\mu\kappa}
 \Big( \delta_{\beta\gamma} \, \zeta_{\mu\kappa}({\bf k}+{\bf G}) \zeta_{\mu\kappa}^*({\bf k}+{\bf G}') \,+ \\ 
 && \qquad \qquad (G_{\beta} - G_{\beta}')  \,
 \zeta_{\mu\kappa,\gamma}({\bf k}+{\bf G}) \zeta_{\mu\kappa}^*({\bf k}+{\bf G}') \Big),
\end{eqnarray}

\noindent where $G=|{\bf G}|$, $v_{\kappa}^{\rm loc}(G)'$ is the first derivative of the spherical atomic pseudopotential, and $\zeta_{\mu\kappa,\gamma}({\bf k}+{\bf G})$ is the ${\bf q}$-derivative along the $\gamma$ direction of the separable nonlocal projector.

\subsection{\label{qderHmet}Second ${\bf q}$-derivative of metric perturbation Hamiltonian}

The first-order Hamiltonian of the metric perturbation is~\cite{metric}
\begin{equation}
\label{Hbeta}
\hat{H}^{(\beta)}_{\bf k,q} =
\hat{T}^{(\beta)}_{\bf k,q} + \hat{V}_{\bf k,q}^{{\rm psp},(\beta)}
 + \hat{V}_{\bf q}^{{\rm H0},(\beta)} + \hat{V}_{\bf q}^{{\rm XC0},(\beta)} + \hat{V}_{\bf q}^{{\rm geom},(\beta)},
\end{equation}
where the terms on the right-hand side correspond to the kinetic ($\hat{T}$), pseudopotential (psp), Hartree (H0),
exchange-correlation (XC0) and geometric contributions to the external potential.
The pseudopotential term, in turn, consists of a local plus a separable contribution,
\begin{equation}
V_{\bf k,q}^{{\rm psp},(\beta)}({\bf G, G}') = V_{\bf q}^{{\rm loc},(\beta)}({\bf G-G}') + V_{\bf k, q}^{{\rm sep},(\beta)}({\bf G,G}')
\end{equation}
The explicit formulas for each of these terms are reported in Ref.~\onlinecite{metric}. 
 In the following, we list the formulas for the second ${\bf q}$-gradients (at ${\bf q}=0$) of these contributions 
 required in the calculation of the \emph{clamped-ion} flexoelectric tensor (see, e.g., Eq.~(\ref{eq_flexo_bks})).


 The kinetic contribution:
\begin{equation}
\begin{split}
 T^{(\beta)}_{{\bf k},\gamma\delta}({\bf G},{\bf G}')=-i \Big(  \delta_{\gamma\delta} (k_{\beta} + G_{\beta}) + 
 \delta_{\beta\gamma} \frac{1}{2}(k_{\delta} + G_{\delta}) + \delta_{\beta\delta} \frac{1}{2}(k_{\gamma} + G_{\gamma}) \Big) \delta_{{\bf G}{\bf G}'}.
\end{split}
\end{equation}

The local part of the pseudopotential: 
\begin{equation}
\begin{split}
 V_{\gamma\delta}^{{\rm loc},\,(\beta)}({\bf G})= 
  -i \frac{1}{\Omega} \sum_{\kappa} e^{- i {\bf G} \tau_{\kappa}} \Bigg( \frac{v_{\kappa}^{\rm loc}(G)'}{G}
  \Big( \delta_{\beta\delta} G_{\gamma}  
  + \delta_{\beta\gamma}G_{\delta} + \delta_{\gamma\delta} G_{\beta} - \frac{G_{\beta} G_{\delta} G_{\gamma}}{G^2} \Big)
  + \frac{v_{\kappa}^{\rm loc}(G)''}{G^2} G_{\beta} G_{\delta} G_{\gamma} \Bigg),
\end{split}
\end{equation}

\noindent with $v_{\kappa}^{\rm loc}(G)''$ being the second derivative of the spherical atomic pseudopotential

The separable part of the pseudopotential:

\begin{equation}
\begin{split}
 & V^{{\rm sep},(\beta)}_{{\bf k},\gamma\delta}= -\frac{i}{\Omega} \sum_{\mu\kappa} e_{\mu\kappa} \, e^{-i({\bf G}-{\bf G}')\tau_{\kappa}}
 \Bigg( \Big( \frac{3}{2} \delta_{\beta\gamma}\, \zeta_{\mu\kappa,\delta}({\bf k}+{\bf G}) + \frac{3}{2} \delta_{\beta\delta} \,\zeta_{\mu\kappa,\gamma}({\bf k}+{\bf G}) +  
 (k_{\beta}+G_{\beta}) \zeta_{\mu\kappa,\delta\gamma}({\bf k}+{\bf G})\Big) \zeta^*_{\mu\kappa}({\bf k} +{\bf G}') \\
 & \quad + \frac{1}{2} \delta_{\beta\delta}\,\zeta_{\mu\kappa}({\bf k}+{\bf G})   \zeta^*_{\mu\kappa,\gamma}({\bf k}+{\bf G}') 
 + \zeta_{\mu\kappa,\delta}({\bf k}+{\bf G})   \zeta^*_{\mu\kappa,\gamma}({\bf k}+{\bf G}') (k_{\beta}+G'_{\beta})
 + \frac{1}{2} \delta_{\beta\gamma} \zeta_{\mu\kappa}({\bf k}+{\bf G})   \zeta^*_{\mu\kappa,\delta}({\bf k}+{\bf G}') \\
 & \quad + \zeta_{\mu\kappa,\gamma}({\bf k}+{\bf G})   \zeta^*_{\mu\kappa,\delta}({\bf k}+{\bf G}') (k_{\beta}+G'_{\beta})  
 + \zeta_{\mu\kappa}({\bf k}+{\bf G})   \zeta^*_{\mu\kappa,\delta\gamma}({\bf k}+{\bf G}') (k_{\beta}+G'_{\beta}) \Bigg),
\end{split}
\end{equation}

\noindent with $\zeta_{\mu\kappa,\gamma\delta}({\bf k}+{\bf G})$ being the second ${\bf q}$-derivative along the $\gamma$ and $\delta$ direction of the separable nonlocal projector.

The remaining terms~\cite{metric} include the XC and geometric contributions that vanish at second order in ${\bf q}$ and a Hartree contribution whose second ${\bf q}$-gradient is,

\begin{equation}
\begin{split}
  \hat{V}_{\delta\gamma}^{{\rm H0},(\beta)}({\bf G})=-i \,8 \pi \frac{n^{(0)}({\bf G})}{G^2} \Big( \frac{4\,G_{\beta}\,G_{\delta}\,G_{\gamma}}{G^4} 
 -\frac{\delta_{\beta\delta}G_{\gamma}+\delta_{\beta\gamma}G_{\delta}+\delta_{\delta\gamma}G_{\beta}}{G^2} \Big),
 \end{split}
\end{equation}

\noindent where $n^{(0)}({\bf G})$ refers to the ground state electron
density.

It is useful, at this point, to perform a further 
rearrangement of $\hat{H}_{{\bf k},\gamma\delta}^{(\beta)}$ by defining
\begin{equation}
\hat{H}_{{\bf k},\gamma}^{(\beta\delta)} = \hat{H}_{{\bf k},\gamma\delta}^{(\beta)} + \hat{H}_{{\bf k},\beta\gamma}^{(\delta)} - 
       \hat{H}_{{\bf k},\beta\delta}^{(\gamma)}.
\end{equation}
This allows us to write the flexoelectric tensor directly in type-II form as 
\begin{equation}
\mu^{\rm II}_{\alpha\gamma,\beta\delta} 
   =  \frac{2}{\Omega} E^{\E_\alpha^* (\beta \delta)}_{\gamma},
\end{equation}
where 
\begin{equation}
\begin{split}
  & E^{\E^*_{\alpha} \, (\beta \delta)}_{\gamma} = s \int_{\rm BZ} [d^3k] \sum_{m}
  E^{\E^*_{\alpha} \, (\beta \delta)}_{m{\bf k},\gamma} 
  + \underbrace{
    \frac{i}{2} \int_{\Omega} \int K_{\gamma}({\bf r},{\bf r}') n^{\E_{\alpha}} ({\bf r})
 n^{\eta_{\beta\delta}}({\bf r}') d^3r d^3r' 
   }_{T_{\rm elst}}, 
\end{split}
\end{equation}
and
\begin{equation}
\begin{split}
  E^{\E^*_{\alpha} \, (\beta \delta)}_{m{\bf k},\gamma} = &
   \underbrace{
   i\langle u_{m{\bf k}}^{\E_{\alpha}} | \partial_{\gamma} \hat{H}^{(0)}_{{\bf k}} | u_{m{\bf k}}^{\eta_{\beta\delta}} \rangle 
   }_{T_1} 
  +  \underbrace{
   i\langle u_{m{\bf k}}^{\E_{\alpha}} | \partial_{\gamma} \hat{Q}_{{\bf k}} \hat{\mathcal{H}}_{{\bf k}}^{\eta_{\beta\delta}} | u_{m{\bf k}}^{(0)} \rangle 
     }_{T_2} +  
    \underbrace{ 
 i\langle u_{m{\bf k}}^{(0)} | \hat{V}^{\E_{\alpha}} \partial_{\gamma} \hat{Q}_{{\bf k}} | u_{m{\bf k}}^{\eta_{\beta\delta}} \rangle 
    }_{T_3} \\ 
 & \quad + \underbrace{
 \frac{1}{2} \langle u_{m{\bf k}}^{\E_{\alpha}} | \hat{H}_{{\bf k},\gamma}^{(\beta \delta)} |  u_{m{\bf k}}^{(0)} \rangle 
   }_{T_4} +  
  \underbrace{
 i \langle i\, u_{m{\bf k},\gamma}^{A_{\alpha}} |  u_{m{\bf k}}^{\eta_{\beta\delta}} \rangle
   }_{T_5}.
\end{split}
\end{equation}

\end{widetext}

\section{\label{g0divs}Treatment of the electrostatic divergence at ${\bf G}=0$}

The local potential diverges at ${\bf G}=0$ because of the Coulomb singularity,~\cite{Gonze}
\begin{equation}
v^{{\rm loc}}_\kappa(q) \sim -\frac{4\pi}{q^2} Z_\kappa,
\label{vq}
\end{equation}
where $Z_\kappa$ is the bare pseudopotential charge.
This means that the ${\bf q}$-derivatives of the local potential contribution to the first-order Hamiltonians
discussed in the previous sections must be calculated with some care regarding the ${\bf G}=0$ component.
To see this, it is useful to rewrite Eq.~(\ref{vq}) as follows,
\begin{equation}
v^{{\rm loc}}_\kappa(q) = \frac{F_\kappa(q)}{q^2},
\end{equation}
where we have introduced the auxiliary function 
\begin{equation}
F_\kappa(q) \sim -4\pi Z_\kappa + \frac{q^2}{2} F_{\kappa}''.
\end{equation} 

Regarding the atomic displacement perturbation, the above definitions lead to the following small-${\bf q}$ 
expansion of the local potential part at ${\bf G}=0$,
\begin{equation}
 V_{{\bf q}}^{\rm loc,\tau_{\kappa \beta}}({\bf G}=0) \sim  -\frac{i q_{\beta}}{\Omega}  \left( -\frac{4\pi Z_\kappa}{q^2} + \frac{F_{\kappa}''}{2}  \right).
\end{equation}
Because of the assumption of short-circuit electrical boundary conditions we shall
drop the divergent term. This leaves us with a constant multiplied by $q_\beta$,
which vanishes in the ${\bf q}\rightarrow 0$ limit. The ${\bf q}$-derivative does not vanish,
\begin{equation}
 V_{\gamma}^{\rm loc,\tau_{\kappa \beta}}({\bf G}=0) \rightarrow  -\frac{i }{2 \Omega} F_{\kappa}'' \delta_{\beta\gamma},
\end{equation}
and we should in principle take it into account in the calculation of the quadrupolar tensor.
However,  in Eq.~(\ref{qdrp_bks}) the operator $\hat{H}_{\bf k \gamma}^{\tau_{\kappa \beta}}$ only appears 
between a conduction-band bra and a valence-band ket. By orthogonality, the above constant contribution is irrelevant
and can be safely discarded.

Regarding the metric perturbation, recall that it vanishes in the ${\bf q}\rightarrow 0$ limit, as
the aforementioned divergence in the local potential contribution exactly cancels with an opposite divergence in
the ``H0'' term.~\cite{metric}
Within the present notation conventions, one has
\begin{equation}
\begin{split}
 & V_{{\bf q}}^{\rm loc+H0,(\beta)}({\bf G}=0) \\ 
 & \quad = -\frac{i}{\Omega}\frac{q_{\beta}}{q^2}\big(\sum_{\kappa}F_{\kappa}(q)-4\pi\Omega n^{(0)}({\bf G}=0)\big) \\
 & \quad = -\frac{i}{2\Omega}q_{\beta}\sum_{\kappa} F_{\kappa}'' + O(q^3),
\end{split}
\end{equation}
where in the last line we have taken into account that $F_{\kappa}(q=0)=4\pi Z_{\kappa}$~\cite{Gonze}, that $n^{(0)}({\bf G}=0)=\frac{\sum_{\kappa} Z_{\kappa}}{\Omega}$ and that odd terms in the Taylor expansion of $F_{\kappa}(q)$ vanish due to the spherical symmetry of the local atomic potentials.
The first ${\bf q}$-derivative of the above yields a well-defined constant,
\begin{equation}
 V_{\gamma}^{\rm loc+H0,(\beta)}({\bf G}=0) = -\frac{i}{2\Omega}\delta_{\beta\gamma}\sum_{\kappa} F_{\kappa}''.
\end{equation}
Recall that the first ${\bf q}$-derivative of the metric-wave perturbation coincides
(modulo a factor of $i$) with the uniform strain perturbation.
Hence, we conclude that to calculate the clamped-ion flexoelectric tensor the ${\bf G}=0$ component of the local potential of the first-order strain Hamiltonian has to be corrected as,

\begin{equation}
V^{{\rm loc},\eta_{\beta\delta}}({\bf G}=0) = V_{\delta}^{{\rm loc},(\beta)}({\bf G}=0)=-\frac{i}{2\Omega}\delta_{\beta\delta}\sum_{\kappa} F_{\kappa}(q)''.
\end{equation}
This correction is important in the calculation of the flexoelectric tensor, since the 
uniform strain operator in Eq.~(\ref{eq_flexo_bks}) appears sandwiched between two 
unperturbed valence states.

\bibliography{merged}

\end{document}